\documentclass[12pt]{article}
\usepackage{amssymb,amsmath,epsfig}
\allowdisplaybreaks
\begin{document}
\title{\bf Effects of Charge on Gravitational Decoupled Anisotropic
Solutions in $f(R)$ Gravity}

\author{M. Sharif \thanks{msharif.math@pu.edu.pk} and Arfa Waseem
\thanks{arfawaseem.pu@gmail.com}\\
Department of Mathematics, University of the Punjab,\\
Quaid-e-Azam Campus, Lahore-54590, Pakistan.}
\date{}

\maketitle

\begin{abstract}
This paper is devoted to studying charged anisotropic static
spherically symmetric solutions through gravitationally decoupled
minimal geometric deformation technique in $f(R)$ gravity. For this
purpose, we first consider the known isotropic Krori-Barua solution
for $f(R)$ Starobinsky model in the interior of a charged stellar
system and then include the effects of two types of anisotropic
solutions. The corresponding field equations are constructed and the
unknown constants are obtained from junction conditions. We analyze
the physical viability and stability of the resulting solutions
through effective energy density, effective radial/tangential
pressure, energy conditions, and causality condition. It is found
that both solutions satisfy the stability range as well as other
physical conditions for specific values of charge as well as model
parameter and anisotropic constant. We conclude that the modified
theory under the influence of charge yields more stable behavior of
the self-gravitating system.
\end{abstract}

\section{Introduction}

Modified theories of gravity have made a remarkable success to
expose invisible effects of dark matter and dark energy. In this
regard, Starobinsky \cite{1} proposed the novel concept of higher
curvature terms to discuss the inflationary scenario in which the
action of general relativity (GR) is modified by $R+\sigma R^{2}$
instead of the Ricci scalar ($R$). The $f(R)$ gravity is the
simplest extension of GR developed by considering an arbitrary
function instead of $R$ in the Einstein-Hilbert action \cite{2}. A
lot of work has been done to discuss the viability as well as
stability of this theory through different approaches
\cite{3}-\cite{10}. The condition of hydrostatic equilibrium in
stellar structure can be employed as a test to analyze the physical
acceptability of $f(R)$ gravity. However, there are some functional
forms of $f(R)$ which do not exhibit the stable stellar system and
are considered unrealistic. During the last few years, many
researchers have discussed stable as well as dynamical unstable
structure of relativistic objects in the framework of this gravity
\cite{11}-\cite{21}.

The study of exact spherical solutions for relativistic objects is a
difficult problem due to the presence of non-linear terms in the
field equations. To resolve this issue, the gravitational decoupling
via minimal geometric deformation (MGD) technique has provided
appreciable results in finding new physically acceptable solutions
for spherical compact configuration. This is a direct, systematic
and simple approach to generate new anisotropic results from perfect
fluid distribution. The decoupling of gravitational sources via MGD
method is a novel concept which displays a large number of
interesting constituents in the construction of new spherical
solutions. This technique was first presented by Ovalle \cite{22} to
obtain new exact solutions of compact stars in the configuration of
braneworld. Later, Ovalle and Linares \cite{23} evaluated the exact
solution for isotropic spherical stellar system and found that their
results show consistency for Tolman-IV solution in the braneworld.

Casadio et al. \cite{24} considered this approach to formulate new
exterior spherical solution which yields singular behavior at
Schwarzschild radius. Ovalle \cite{25} formulated exact anisotropic
spherical solutions from perfect fluid by decoupling the
gravitational source through this technique. Ovalle et al. \cite{26}
extended interior isotropic solution by including the anisotropic
source term for spherical stellar object and examined the graphical
description of effective radial pressure, anisotropic factor and
gravitational redshift. In the same scenario, Gabbanelli et al.
\cite{27} obtained physically viable anisotropic solutions by taking
isotropic Durgapal-Fuloria stellar system and discussed the
graphical interpretation of matter variables. Graterol \cite{28}
deformed the Buchdahl solution to attain analytic solution and
calculated the unknown constants in new anisotropic solution via
matching conditions. Recently, Panotopoulos and Rinc\'{o}n
\cite{28a} found exact analytical solutions in 3-dimensional gravity
using the MGD approach in a cloud of strings and analyzed the
behavior of matter variables graphically.

The presence of electric field in self-gravitating object has
significant importance in describing their evolution and stable
structure. In the analysis of astrophysical scenarios, it is
observed that a star needs a large amount of charge to repel strong
gravitational pull. In the context of GR as well as alternative
theories, a numerous research has been done to examine the influence
of charge on the physical behavior of celestial bodies. It is found
that the existence of electric field leads to more stable
configuration of stellar systems \cite{29}-\cite{35}. Recently,
Sharif and Sadiq \cite{36} explored charged anisotropic spherical
solutions through MGD technique by considering Krori-Barua solution
as a known charged isotropic solution. They observed the role of
physical parameters, stability and energy conditions for different
values of charge parameter to analyze the regularity of their
solutions.

In this paper, we discuss the influence of charge as well as
modified theory on new exact anisotropic spherically symmetric
solutions through MGD approach. We consider the well-known
Krori-Barua solution as the known charged isotropic solution and
analyze the analytic anisotropic solutions in $f(R)$ gravity. The
paper is arranged in the following pattern. The next section
provides some basics of $f(R)$ field equations with electromagnetic
field corresponding to a realistic model and uses MGD method to gain
two sets of decoupled equations. Section \textbf{3} deals with the
Krori-Barua charged isotropic solution whose unknown constants are
found through matching conditions. We obtain two new charged
anisotropic solutions whose viability is investigated via graphical
analysis. Finally, we present our conclusive remarks in the last
section.

\section{Gravitational Decoupled Field Equations}

The modification of the Einstein-Hilbert action in the presence of
matter Lagrangian ($\mathcal{L}_{m}$) depending upon the metric
tensor ($g_{\xi\eta}$) is presented by the action \cite{2}
\begin{equation}\label{1}
\mathcal{I}_{f(R)}=\int d^{4}x\sqrt{-g}
\left[\frac{f(R)}{2\kappa}+\mathcal{L}_{m}\right],
\end{equation}
where $f(R)$ represents an arbitrary function of the curvature
scalar and $\kappa=1$ indicates the coupling constant. The field
equations corresponding to action (\ref{1}) are
\begin{equation}\label{2}
f_{R}R_{\xi\eta}-\frac{1}{2}g_{\xi\eta}f(R)-(\nabla_{\xi}
\nabla_{\eta}-g_{\xi\eta}\Box)f_{R}=T_{\xi\eta}^{(m)} +E_{\xi\eta},
\end{equation}
where $f_{R}=\frac{\partial f}{\partial R}$,
$\Box=g^{\xi\eta}\nabla_{\xi}\nabla_{\eta}$ and $\nabla_{\xi}$
stands for covariant derivative. An alternate expression of
Eq.(\ref{2}) can be expressed as
\begin{equation}\label{3}
G_{\xi\eta}=R_{\xi\eta}-\frac{1}{2}Rg_{\xi\eta}=\frac{1}{f_{R}}
T_{\xi\eta}^{(tot)},
\end{equation}
where
\begin{equation}\nonumber
T_{\xi\eta}^{(tot)}=T_{\xi\eta}^{(m)}+E_{\xi\eta}+\chi\Theta_{\xi\eta}
+T_{\xi\eta}^{(D)}.
\end{equation}
Here, $T_{\xi\eta}^{(m)}$ represents the standard energy-momentum
tensor whose mathematical form corresponding to perfect fluid
distribution comprising the four-velocity field $(U_{\xi})$, energy
density $(\rho)$ and pressure $(p)$ is as follows
\begin{equation}\label{4}
T_{\xi\eta}^{(m)}=(\rho+p)U_{\xi}U_{\eta}-pg_{\xi\eta},
\end{equation}
$E_{\xi\eta}$ denotes the electromagnetic energy-momentum tensor
defined as
\begin{equation}\label{5}
E_{\xi\eta}=\frac{1}{4\pi}\left(\frac{F^{\alpha\beta}
F_{\alpha\beta}g_{\xi\eta}}{4}-F^{\alpha}_{\xi}F_{\eta\alpha}\right),
\end{equation}
$T_{\xi\eta}^{(D)}$ shows the constituents that appear from the
contribution of modified terms in the energy-momentum tensor given
by
\begin{equation}\nonumber
T_{\xi\eta}^{(D)}=\left(\frac{f(R)-Rf_{R}}{2}\right)g_{\xi\eta}
+(\nabla_{\xi}\nabla_{\eta}-g_{\xi\eta}\Box)f_{R},
\end{equation}
and $\Theta_{\xi\eta}$ illustrates an extra term that is
gravitationally coupled via constant $\chi$ which may comprise new
fields (such as scalar, vector and tensor) and may induce anisotropy
in relativistic objects \cite{37}.

In order to portray the internal configuration of self-gravitating
objects, we consider a static spherically symmetric spacetime
\begin{equation}\label{6}
ds^{2}_{-}=e^{\mu(r)}dt^{2}-e^{\lambda(r)}dr^{2}
-r^{2}(d\theta^{2}+\sin^{2}\theta d\phi^{2}),
\end{equation}
where the metric potentials ($\mu$ and $\lambda$) depend only on
radial coordinate $r$ ranging from the center $(r=0)$ to the surface
$(r=\mathcal{R})$ of star while the four-velocity yields
$U^{\xi}=e^{-\frac{\mu}{2}}\delta^{\xi}_{0}$ for $0\leq r\leq
\mathcal{R}$. The Maxwell field equations are
\begin{equation}\nonumber
F_{\xi\eta;\alpha}=0,\quad F^{\xi\eta}_{~~;\eta}=4\pi j^{\xi},
\end{equation}
where $j^{\xi}$ is the four current density. Here, we take a
comoving frame in which the charge parameter remains at rest and
consequently, no more magnetic field is produced. The four current
density and four potential in comoving coordinates obey the
following identities
\begin{equation}\label{7}
j^{\xi}=\zeta(r)U^{\xi},
\quad\varphi^{\xi}=\varphi(r)\delta^{\xi}_{0},
\end{equation}
where $\zeta$ indicates the charge density. The Maxwell field
equations corresponding to the metric (\ref{6}) become
\begin{eqnarray}\label{8}
\varphi''-\left(\frac{\mu'+\lambda'}{2}-\frac{2}{r}\right)
\varphi'=4\pi\zeta(r)e^{\frac{\mu}{2}+\lambda},
\end{eqnarray}
where prime reveals derivative with respect to radial coordinate.
Integration of the above equation leads to
\begin{equation}\label{9}
\varphi'=\frac{qe^{\frac{\mu+\lambda}{2}}}{r^{2}},
\end{equation}
where $q$ represents charge in the interior region of star. The
field equations of $f(R)$ gravity (\ref{3}) for spherically
symmetric spacetime are
\begin{eqnarray}\nonumber
\frac{1}{r^{2}}+e^{-\lambda}\left(\frac{\lambda'}{r}-\frac{1}
{r^{2}}\right)&=&\frac{1}{f_{R}}\left[\rho+\chi\Theta_{0}^{0}
+\frac{q^{2}}{8\pi r^{4}}+\frac{f-Rf_{R}}{2}\right.
\\\label{10}&-&\left.e^{-\lambda}\left\{\left(\frac{\lambda'}{2}
-\frac{2}{r}\right)f_{R}'-f_{R}''\right\}\right],
\\\nonumber
e^{-\lambda}\left(\frac{\mu'}{r}+\frac{1}{r^{2}}\right)
-\frac{1}{r^{2}}&=&\frac{1}{f_{R}}\left[p-\chi\Theta_{1}^{1}
-\frac{q^{2}}{8\pi r^{4}}-\frac{f-Rf_{R}}{2}\right.
\\\label{11}&-&\left.e^{-\lambda}\left(\frac{\mu'}{2}
+\frac{2}{r}\right)f_{R}'\right],
\\\nonumber
e^{-\lambda}\left(\frac{\mu''}{2}+\frac{\mu'}{2r}
-\frac{\lambda'}{2r}+\frac{\mu'^{2}}{4}-\frac{\mu'\lambda'}{4}\right)
&=&\frac{1}{f_{R}}\left[p-\chi\Theta_{2}^{2}+\frac{q^{2}}{8\pi
r^{4}}-\frac{f-Rf_{R}}{2}\right.\\\nonumber&+&\left.e^{-\lambda}
\left\{\left(\frac{\lambda'}{2}-\frac{\mu'}{2}
-\frac{1}{r}\right)f_{R}'-f_{R}''\right\}\right].\\\label{12}
\end{eqnarray}
In $f(R)$ gravity, the conservation equation
$(\nabla_{\xi}T^{\xi\eta(tot)}=0)$ is also satisfied whose
expression corresponding to (\ref{6}) becomes
\begin{equation}\label{13}
p'+\frac{\mu'}{2}(\rho+p)-\chi(\Theta_{1}^{1})'+\frac{\mu'\chi}{2}
(\Theta_{0}^{0}-\Theta_{1}^{1})+\frac{2\chi}{r}(\Theta_{2}^{2}
-\Theta_{1}^{1})-\frac{qq'}{4\pi r^{4}}=0.
\end{equation}
Notice that the standard conservation equation for charged perfect
fluid configuration can be recovered for $\chi=0$.

In the analysis of early universe, various inflationary models are
developed on scalar fields originating from super-string and
super-gravity theories. Starobinsky \cite{1} suggested the first
inflation model which corresponds to the conformal deviation in
quantum gravity given by
\begin{equation}\label{14}
f(R)=R+\sigma R^{2},
\end{equation}
where $\sigma\in\mathbb{R}$. It is observed that this functional
form may lead to the accelerated cosmic expansion due to the impact
of $\sigma R^{2}$ term. This model is also found to be consistent
with the temperature anisotropies detected in cosmic microwave
background and hence can be served as a reliable alternative for the
inflationary candidates \cite{38}. The signature of $f_{RR}$ is of
fundamental importance as it examines how much this modified theory
reaches to the GR limit. The consistency of this model is attained
for $\sigma>0$ which is directly related to $f_{RR}>0$. In the
analysis of self-gravitating objects, Zubair and Abbas \cite{21}
established that the acceptable values of $\sigma$ lie in the range
$0 < \sigma < 6$. The results of GR can be regained from the
proposed model for $\sigma=0$. This model has extensively been
implemented in literature to narrate various cosmological issues.

The field equations (\ref{10})-(\ref{12}) corresponding to the model
(\ref{14}) become
\begin{eqnarray}\label{15}
\rho+\sigma F_{1}+\frac{q^{2}}{8\pi r^{4}}
+\chi\Theta_{0}^{0}&=&\frac{1}{r^{2}}-e^{-\lambda}
\left(\frac{1}{r^{2}}-\frac{\lambda'}{r}\right),
\\\label{16}
p+\sigma F_{2}-\frac{q^{2}}{8\pi r^{4}}
-\chi\Theta_{1}^{1}&=&e^{-\lambda}\left(\frac{\mu'}{r}
+\frac{1}{r^{2}}\right)-\frac{1}{r^{2}},
\\\label{17}
p+\sigma F_{3}+\frac{q^{2}}{8\pi r^{4}}-\chi\Theta_{2}^{2}&=&
e^{-\lambda} \left(\frac{\mu''} {2}-\frac{\lambda'}{2r}
+\frac{\mu'}{2r}-\frac{\mu'\lambda'}{4} +\frac{\mu'^{2}}{4}\right),
\end{eqnarray}
where $F_{1}$, $F_{2}$ and $F_{3}$ are of the following forms
\begin{eqnarray}\nonumber
F_{1}&=&-\frac{R^{2}}{2}+2e^{-\lambda} \left\{R''
-\left(\frac{\lambda'}{2}-\frac{2}{r}\right)R'\right\}
-2R\left\{\frac{1}{r^{2}}+e^{-\lambda}\left(\frac{\lambda'}{r}
-\frac{1}{r^{2}}\right)\right\},\\\nonumber
F_{2}&=&\frac{R^{2}}{2}-2e^{-\lambda}\left(\frac{\mu'}
{2}+\frac{2}{r}\right)R'-2R\left\{e^{-\lambda}
\left(\frac{\mu'}{r}+\frac{1}{r^{2}}\right)
-\frac{1}{r^{2}}\right\},\\\nonumber
F_{3}&=&\frac{R^{2}}{2}+2e^{-\lambda}\left\{\left(\frac{\lambda'
-\mu'}{2}-\frac{1}{r}\right)R'-\left(\frac{\mu''}{2}
-\frac{\lambda'-\mu'}{2r}+\frac{\mu'^{2}}{4}
-\frac{\mu'\lambda'}{4}\right)R\right.
\\\nonumber&-&\left.R''\right\}.
\end{eqnarray}
Here, we have a system of non-linear differential equations
(\ref{13}) and (\ref{15})-(\ref{17}) which contains eight unknown
functions ($\mu$, $\lambda$, $\rho$, $p$, $q$, $\Theta_{0}^{0}$,
$\Theta_{1}^{1}$ and $\Theta_{2}^{2}$). In order to close the
system, we employ a systematic approach proposed by Ovalle
\cite{26}. For the set of equations (\ref{15})-(\ref{17}), the
matter variables (effective energy density, effective radial and
effective tangential pressures) can be identified as
\begin{equation}\label{18}
\tilde{\rho}^{eff}=\rho^{eff}+\chi\Theta_{0}^{0},\quad
\tilde{p}_{r}^{eff}=p^{eff}-\chi\Theta_{1}^{1},\quad
\tilde{p}_{t}^{eff}=p^{eff}-\chi\Theta_{2}^{2},
\end{equation}
where $\rho^{eff}$ and $p^{eff}$ denote $\rho+\sigma F_{1}$ and
$p+\sigma F_{2}$, respectively. From these definitions, it is
clearly observed that the source $\Theta_{\xi\eta}$ generates
anisotropy in the interior of self-gravitating systems. The
effective anisotropic factor is defined as follows
\begin{equation}\label{19}
\tilde{\Delta}^{eff}=\tilde{p}_{t}^{eff}-\tilde{p}_{r}^{eff}
=\chi(\Theta_{1}^{1}-\Theta_{2}^{2}).
\end{equation}
It is worth mentioning here that the anisotropy factor vanishes for
$\chi=0$.

\subsection{The MGD Approach}

In this section, we consider a new technique known as gravitational
decoupling through MGD approach to solve a set of non-linear
differential equations (\ref{15})-(\ref{17}). This technique is used
to transform the field equations in such a way that the source
$\Theta_{\xi\eta}$ provides the form of effective equations which
may produce an anisotropy. The most fundamental constituent of this
technique is the perfect fluid solution ($\alpha$, $\nu$, $\rho$,
$p$ and $q$) with the metric
\begin{equation}\label{20}
ds^{2}=e^{\alpha(r)}dt^{2}-\frac{dr^{2}}{\nu(r)}-r^{2}(d\theta^{2}
+\sin^{2}\theta d\phi^{2}),
\end{equation}
where $\nu(r)=1-\frac{2m(r)}{r}+\frac{q^{2}}{r^{2}}$ is the usual GR
expression that contains the Misner-Sharp mass $m$ and charge $q$.
The influence of source $\Theta_{\xi\eta}$ in charged isotropic
model can be encoded by the implementation of geometric deformation
on the metric potentials ($\alpha$ and $\nu$) through a linear
mapping defined as
\begin{equation}\label{21}
\alpha\mapsto\mu=\alpha+\chi g, \quad \nu\mapsto
e^{-\lambda}=\nu+\chi I,
\end{equation}
where $g$ and $I$ are the corresponding deformations offered to
temporal and radial metric ingredients, respectively. It is
worthwhile to mention here that the geometric deformations in
(\ref{21}) are entirely radial functions which confirm the spherical
symmetry of the solution. Among these deformations, MGD corresponds
to
\begin{equation}\label{22}
g\mapsto 0, \quad I\mapsto I^{*},
\end{equation}
where $I^{*}$ shows the minimal geometric deformation. In this case,
the deformation is applied only on the radial component whereas the
temporal one remains unchanged. Thus, the anisotropic source
$\Theta_{\xi\eta}$ is purely merged in the radial deformation
denoted by
\begin{equation}\label{23}
\alpha\mapsto\mu=\alpha, \quad \nu\mapsto e^{-\lambda}=\nu+\chi
I^{*}.
\end{equation}
Inserting Eq.(\ref{23}) into Eqs.(\ref{15})-(\ref{17}), the system
decouples into two sets.

The first set corresponds to $\chi=0$ leading to the following
charged perfect fluid matter configuration
\begin{eqnarray}\label{24}
\rho^{eff}+\frac{q^{2}}{8\pi r^{4}}&=&\rho+\sigma
F_{1}+\frac{q^{2}}{8\pi r^{4}}=\frac{1}{r^{2}}-\frac{\nu}{r^{2}}
-\frac{\nu'}{r},\\\label{25} p^{eff}-\frac{q^{2}}{8\pi
r^{4}}&=&p+\sigma F_{2}-\frac{q^{2}}{8\pi r^{4}}
=-\frac{1}{r^{2}}+\frac{\nu}{r}\left(\frac{1}{r}+\mu'
\right),\\\nonumber p^{eff}+\frac{q^{2}}{8\pi r^{4}}&=&p+\sigma
F_{3}+\frac{q^{2}}{8\pi r^{4}}=\frac{\nu}{4}
\left(2\mu''+\mu'^{2}+\frac{2\mu'}{r}\right)
+\frac{\nu'}{4}\left(\mu'+\frac{2}{r}\right),\\\label{26}
\end{eqnarray}
as well as the conservation equation
\begin{equation}\label{27}
p'+\frac{\mu'}{2}(\rho+p)-\frac{qq'}{4\pi r^{4}}=0.
\end{equation}
The second set of equations comprising the source $\Theta_{\xi\eta}$
yields
\begin{eqnarray}\label{28}
\Theta_{0}^{0}&=&-\frac{1}{r}(I^{*'}+\frac{I^{*}}{r}),\\\label{29}
\Theta_{1}^{1}&=& =-\frac{I^{*}}{r}\left(\frac{1}{r}+\mu'
\right),\\\label{30} \Theta_{2}^{2}&=&-\frac{I^{*}}{4}
\left(2\mu''+\mu'^{2}+\frac{2\mu'}{r}\right)
-\frac{I^{*'}}{4}\left(\mu'+\frac{2}{r}\right),
\end{eqnarray}
and the conservation equation, $\nabla_{\xi}\Theta_{\xi\eta}=0$, is
explicitly expressed as
\begin{equation}\label{31}
(\Theta_{1}^{1})'-\frac{\mu'}{2}
(\Theta_{0}^{0}-\Theta_{1}^{1})-\frac{2}{r}(\Theta_{2}^{2}
-\Theta_{1}^{1})=0.
\end{equation}
From Eqs.(\ref{27}) and (\ref{31}), it is clearly shown that there
is no change of energy-momentum tensor between the charged perfect
fluid distribution and the source $\Theta_{\xi\eta}$ which assures
that their interaction is absolutely gravitational. It is noted that
the set of Eqs.(\ref{28})-(\ref{30}) are similar to the spherically
symmetric field equations for anisotropic matter distribution with
source $\Theta_{\xi\eta}$ relative to the metric
\begin{equation}\label{32}
ds^{2}=e^{\mu(r)}dt^{2}-\frac{dr^{2}}{I^{*}(r)}
-r^{2}(d\theta^{2}+\sin^{2}\theta d\phi^{2}).
\end{equation}
However, the expressions on right-hand side of
Eqs.(\ref{28})-(\ref{30}) are not the standard one as they show
deviation from anisotropic solution by the factor $\frac{1}{r^{2}}$
which represent the matter constituents as
\begin{eqnarray}\label{33}
\tilde{\rho}^{eff}+\frac{q^{2}}{8\pi
r^{4}}&=&\Theta_{0}^{0*}=\Theta_{0}^{0}+\frac{1}{r^{2}},
\\\label{34}
\tilde{p}_{r}^{eff}-\frac{q^{2}}{8\pi
r^{4}}&=&\Theta_{1}^{1*}=\Theta_{1}^{1}+\frac{1}{r^{2}},
\\\label{35}
\tilde{p}_{t}^{eff}+\frac{q^{2}}{8\pi
r^{4}}&=&\Theta_{2}^{2*}=\Theta_{2}^{2}=\Theta_{3}^{3*}
=\Theta_{3}^{3}.
\end{eqnarray}
Thus, the MGD approach has turned the indefinite system
(\ref{15})-(\ref{17}) into a set of equations for charged perfect
fluid along with a set of four unknown functions
$(I^{*},\Theta_{0}^{0},\Theta_{1}^{1},\Theta_{2}^{2})$ satisfying
the anisotropic system (\ref{33})-(\ref{35}). Hence, the system
(\ref{15})-(\ref{17}) has been decoupled successfully.

\subsection{Junction Conditions}

In the evolution of self-gravitating systems, the junction
conditions has a dynamical contribution that provide a linear
relation between interior as well as exterior metrics at the
boundary of star to analyze the physical behavior of stellar
objects. In this work, the interior geometry of stellar distribution
is obtained through MGD as
\begin{equation}\label{36}
ds^{2}=e^{\mu_{-}(r)}dt^{2}-\left(1-\frac{2\tilde{m}(r)}{r}
+\frac{q^{2}}{r^{2}}\right)^{-1}dr^{2}-r^{2}(d\theta^{2}
+\sin^{2}\theta d\phi^{2}),
\end{equation}
where the internal mass function is $\tilde{m}=m(r)-\frac{\chi r}{2}
I^{*}(r)$. For a smooth relation between the geometries (interior
and exterior) of star, the general exterior metric is
\begin{equation}\label{37}
ds^{2}=e^{\mu_{+}(r)}dt^{2}-e^{\lambda_{+}(r)}dr^{2}-r^{2}
(d\theta^{2}+\sin^{2}\theta d\phi^{2}).
\end{equation}
The continuity of the first fundamental form of junction conditions
over the hypersurface $(\Sigma)$ leads to $[ds^{2}]_{\Sigma}=0$,
where $[S]_{\Sigma}\equiv S^{+}(\mathcal{R})-S^{-}(\mathcal{R})$ for
any function $S=S(r)$ gives
\begin{eqnarray}\label{38}
\mu_{+}(\mathcal{R})=\mu_{-}(\mathcal{R}),\quad
1-\frac{2M_{0}}{\mathcal{R}}+\frac{Q_{o}^{2}}{R^{2}}+\chi
I^{*}(\mathcal{R})=e^{-\lambda_{+}(\mathcal{R})}.
\end{eqnarray}
Here, $M_{0}=m(\mathcal{R})$, $Q_{0}$ and $I^{*}(\mathcal{R})$
denote the total mass, total charge and deformation at the boundary
of star, respectively.

The continuity of the second fundamental form
($[T_{\xi\eta}^{(tot)}K^{\eta}]_{\Sigma}=0$ with $K^{\eta}$ as a
unit four-vector in radial direction) yields \cite{26}
\begin{equation}\nonumber
p^{eff}(\mathcal{R})-\frac{Q_{0}^{2}}{8\pi \mathcal{R}^{4}}-\chi
(\Theta_{1}^{1}(\mathcal{R}))^{-}=-\chi
(\Theta_{1}^{1}(\mathcal{R}))^{+},
\end{equation}
which gives rise to
\begin{equation}\label{39}
p^{eff}(\mathcal{R})-\frac{Q_{0}^{2}}{8\pi\mathcal{R}^{4}}
+\frac{\chi I^{*}(\mathcal{R})}{\mathcal{R}}
\left(\frac{1}{\mathcal{R}}+\mu'(\mathcal{R})\right)=\frac{\chi
h^{*}(\mathcal{R})}{\mathcal{R}^{2}}\left(1+\frac{2M\mathcal{R}
-2\mathcal{Q}^{2}}{\mathcal{R}^{2}-2M\mathcal{R}+\mathcal{Q}^{2}}\right),
\end{equation}
where $M$, $\mathcal{Q}$ show the mass as well as charge of exterior
geometry and $h^{*}$ denotes the outer radial geometric deformation
for Riessner-Nordstr\"{o}m (RN) metric in the presence of source
$\Theta_{\xi\eta}$ described by
\begin{equation}\label{40}
ds^{2}=\left(1-\frac{2M}{r}+\frac{\mathcal{Q}^{2}}{r^{2}}\right)dt^{2}
-\frac{dr^{2}}{\left(1-\frac{2M}{r}+\frac{\mathcal{Q}^{2}}{r^{2}}+\chi
h^{*}\right)}-r^{2}(d\theta^{2}+\sin^{2}\theta d\phi^{2}).
\end{equation}
The necessary and sufficient conditions for a direct relation
between MGD interior and RN exterior metrics (filled with source
$\Theta_{\xi\eta}$) are provided by the constraints (\ref{38}) and
(\ref{39}). If we consider the exterior spacetime as the standard RN
metric ($h^{*}=0$), then
\begin{equation}\label{41}
\tilde{p}^{eff}(\mathcal{R})-\frac{Q_{0}^{2}}{8\pi
\mathcal{R}^{4}}\equiv p^{eff}(\mathcal{R})-\frac{Q_{0}^{2}}{8\pi
\mathcal{R}^{4}}+\frac{\chi
I^{*}}{\mathcal{R}}\left(\frac{1}{\mathcal{R}}+\mu'\right)=0.
\end{equation}

\section{Anisotropic Solutions}

In order to attain the anisotropic solutions for charged stellar
system through MGD technique, we need solution of the field
equations for charged perfect fluid spherical system in $f(R)$
gravity. In this regard, we consider the Krori-Barua solution that
has become a subject of great interest due to its singularity free
nature \cite{39}. This solution has attained much attention in
analyzing the behavior of charged stellar systems both in GR as well
as modified theories. In the background of $f(R)$ gravity, Momeni et
al. \cite{20} examined the stellar configuration for this solution
without charge by employing extended forms of
Tolman-Oppenheimer-Volkoff equations. In the same theory, Zubair and
Abbas \cite{21} used this solution to investigate physical
characteristics as well as stable structure of anisotropic compact
objects.

The Krori-Barua solution yields a consistent as well as realistic
method in the analysis of stellar evolution. For charged perfect
fluid distribution in $f(R)$ gravity, this solution is defined as
\begin{eqnarray}\label{42}
e^{\mu(r)}&=&e^{\mathcal{B}r^{2}+\mathcal{C}},\\\label{43}
e^{\lambda(r)}&=&\nu^{-1}=e^{\mathcal{A}r^{2}},\\\nonumber
\rho^{eff}&=&\frac{1}{2r^{2}}+\frac{e^{-\mathcal{A}r^{2}}}{2}
\left(5\mathcal{A}-\frac{1}{r^{2}}-\mathcal{B}^{2}r^{2}
+\mathcal{AB}r^{2}\right)+\frac{6\sigma e^{-2\mathcal{A}r^{2}}}
{r^{4}}\left[-4+\mathcal{B}r^{2}(1\right.\\\nonumber&-&\left.
\mathcal{B}r^{2}+\mathcal{B}^{2}r^{4})+e^{\mathcal{A}r^{2}}(4
+\mathcal{A}r^{2}-\mathcal{B}r^{2})+6\mathcal{A}^{3}r^{6}(2
+\mathcal{B}r^{2})-\mathcal{A}^{2}r^{4}(16\right.\\\label{44}&+&\left.
33\mathcal{B}r^{2}+6\mathcal{B}^{2}r^{4})+\mathcal{A}r^{2}
(-5+17\mathcal{B}r^{2}+14\mathcal{B}^{2}r^{4})\right],\\\nonumber
p^{eff}&=&\frac{-1}{2r^{2}}+\frac{e^{-\mathcal{A}r^{2}}}{2}
\left(4\mathcal{B}+\frac{1}{r^{2}}-\mathcal{A}-\mathcal{AB}r^{2}
+\mathcal{B}^{2}r^{2}\right)-\frac{2\sigma e^{-2\mathcal{A}r^{2}}}
{r^{2}}\left[6\mathcal{A}^{3}r^{4}(2\right.\\\nonumber&+&\left.
\mathcal{B}r^{2})-\mathcal{A}^{2}r^{2}(20+43\mathcal{B}r^{2}
+10\mathcal{B}^{2}r^{4})+\mathcal{A}(-3+3e^{\mathcal{A}r^{2}}
+35\mathcal{B}r^{2}+34\right.\\\label{45}&\times&\left.\mathcal{B}^{2}
r^{4}+4\mathcal{B}^{3}r^{6})-\mathcal{B}(-3+11\mathcal{B}r^{2}
+5\mathcal{B}^{2}r^{4}+3e^{\mathcal{A}r^{2}})\right],\\\nonumber
q^{2}&=&4\pi r^{4}\left[\frac{1}{r^{2}}+{e^{-\mathcal{A}r^{2}}}
\left(\mathcal{B}^{2}r^{2}-\frac{1}{r^{2}}-\mathcal{A}
-\mathcal{AB}r^{2}\right)-\frac{4\sigma e^{-2\mathcal{A}r^{2}}}
{r^{4}}\left\{e^{2\mathcal{A}r^{2}}-7\right.\right.\\\nonumber&+&
\left.\left.\mathcal{B}r^{2}(3-3\mathcal{B}^{2}r^{4}
-\mathcal{B}^{3}r^{6})+6\mathcal{A}^{3}r^{6}(2+\mathcal{B}r^{2})
-\mathcal{A}^{2}r^{4}(8+31\mathcal{B}r^{2} +7\mathcal{B}^{2}
\right.\right.\\\nonumber&\times&\left.\left.r^{4})+\mathcal{A}r^{2}
(-11+3\mathcal{B}r^{2}+16\mathcal{B}^{2}r^{4}+2\mathcal{B}^{3}
r^{6})+3\mathcal{B}(-3+11\mathcal{B}r^{2}+5\mathcal{B}^{2}
\right.\right.\\\label{46}&\times&\left.\left.r^{4})+3e^{\mathcal{A}
r^{2}}(2+\mathcal{A}r^{2}-\mathcal{B}r^{2})\right\}\right],
\end{eqnarray}
where the triplet ($\mathcal{A}$, $\mathcal{B}$, $\mathcal{C}$)
represents unknown constants that can be computed from matching
conditions. From the matching between interior and exterior
geometries of stellar object, the continuity of metric variables
$g_{tt}$, $g_{rr}$ and $g_{tt,r}$ leads to the following forms of
$\mathcal{A}$, $\mathcal{B}$ and $\mathcal{C}$
\begin{eqnarray}\label{47}
\mathcal{A}&=&\frac{-1}{\mathcal{R}^{2}}\ln\left(1-\frac{2M_{0}}
{\mathcal{R}}+\frac{Q_{0}^{2}}{\mathcal{R}^{2}}\right),
\\\label{48}\mathcal{B}&=&\frac{1}{\mathcal{R}^{2}}\left(\frac{M_{0}}
{\mathcal{R}}-\frac{Q_{0}^{2}}{\mathcal{R}^{2}}\right)\left(1
-\frac{2M_{0}}{\mathcal{R}}+\frac{Q_{0}^{2}}{\mathcal{R}^{2}}\right)^{-1},
\\\label{49}\mathcal{C}&=&\ln\left(1-\frac{2M_{0}}{\mathcal{R}}
+\frac{Q_{0}^{2}}{\mathcal{R}^{2}}\right)-\frac{M_{0}\mathcal{R}
-Q_{0}^{2}}{\mathcal{R}^{2}-2M_{0}\mathcal{R}+Q_{0}^{2}},
\end{eqnarray}
along with the compactness factor
$\frac{2M_{0}}{\mathcal{R}}<\frac{8}{9}$. These expressions assure
the continuity of charged isotropic solution (\ref{42})-(\ref{46})
with the exterior RN geometry at star's surface which will surely be
changed with the presence of source $\Theta_{\xi\eta}$ in the
interior region.

In order to have anisotropic solution, i.e., for $\chi\neq0$ in the
interior of spherical stellar object, the radial as well as temporal
metric constituents are given by Eqs.(\ref{23}) and (\ref{42}),
respectively. The geometric deformation $(I^{*})$ and the source
term ($\Theta_{\xi\eta}$) are related through
Eqs.(\ref{28})-(\ref{30}) whose solution will be evaluated by
considering some additional constraints. For this purpose, we impose
some conditions to derive two physically consistent interior
solutions in the following subsections.

\subsection{The First Solution}

In this section, we adopt a constraint on $\Theta_{1}^{1}$ and find
a solution of the field equations for the source term
$\Theta_{\xi\eta}$ and deformation function $I^{*}$. From
Eq.(\ref{41}), it is observed that RN exterior geometry shows
compatibility with the isotropic interior metric as long as
$p^{eff}(\mathcal{R})-\frac{Q_{0}^{2}}{8\pi
\mathcal{R}^{4}}\sim\chi(\Theta_{1}^{1}(\mathcal{R}))_{-}$. In order
to satisfy this requirement, the direct choice is to consider
\cite{36}
\begin{equation}\label{50}
\Theta_{1}^{1}=p^{eff}-\frac{q^{2}}{8\pi r^{4}}.
\end{equation}
With the help of Eqs.(\ref{25}) and (\ref{29}), this leads to
\begin{equation}\label{51}
I^{*}=\frac{1}{1+\mu'r}-\nu,
\end{equation}
which yields the radial metric coefficient as
\begin{equation}\label{52}
e^{-\lambda}=(1-\chi)\nu+\frac{\chi}{1+2\mathcal{B}r^{2}}.
\end{equation}
The metric constituents of interior geometry in Eqs.(\ref{42}) and
(\ref{52}) illustrate the minimally deformed Krori-Barua solution
through the generic anisotropic source $\Theta_{\xi\eta}$. For
$\chi\rightarrow0$, Eq.(\ref{52}) reduces to the standard spherical
solution.

Now, employing junction conditions, the continuity of first
fundamental form leads to
\begin{equation}\label{53}
\ln\left(1-\frac{2M}{\mathcal{R}}+\frac{\mathcal{Q}^{2}}
{\mathcal{R}^{2}}\right)=\mathcal{BR}^{2}+\mathcal{C},
\end{equation}
which further gives rise to
\begin{equation}\label{54}
1-\frac{2M}{\mathcal{R}}+\frac{\mathcal{Q}^{2}}{R^{2}}=(1
-\chi)\nu+\frac{\chi}{1+2\mathcal{B}\mathcal{R}^{2}}.
\end{equation}
Similarly, the continuity of second fundamental form
($p^{eff}-\frac{Q_{0}^{2}}{8\pi\mathcal{R}^{4}}+\chi(\Theta_{1}^{1}
(\mathcal{R}))_{-}\\=0$) along with Eq.(\ref{50}) provides
\begin{equation}\label{55}
p^{eff}(\mathcal{R})-\frac{Q_{0}^{2}}{8\pi\mathcal{R}^{4}}=0\quad\Rightarrow\quad
\mathcal{A}=\frac{\ln(1+2\mathcal{B}\mathcal{R}^{2})}{\mathcal{R}^{2}}.
\end{equation}
To evaluate the expression of mass, Eq.(\ref{54}) leads to
\begin{equation}\label{56}
\frac{2M}{\mathcal{R}}=\frac{2M_{0}}{\mathcal{R}}+\frac{\mathcal{Q}^{2}
-Q_{0}^{2}}{\mathcal{R}^{2}}+\chi\left(1 -\frac{2M_{0}}{\mathcal{R}}
+\frac{Q_{0}^{2}}{\mathcal{R}^{2}}-\frac{1}{1+2B\mathcal{R}^{2}}\right).
\end{equation}
On inserting the above expression in Eq.(\ref{53}), we obtain
\begin{equation}\label{57}
\mathcal{B}\mathcal{R}^{2}+\mathcal{C}=\ln\left[\left(1-\frac{2M_{0}}
{\mathcal{R}}+\frac{Q_{0}^{2}}{\mathcal{R}^{2}}\right)(1-\chi)
+\frac{\chi}{1+2\mathcal{B}\mathcal{R}^{2}}\right],
\end{equation}
where the constant $\mathcal{C}$ can be described in terms of
$\mathcal{B}$. The set of equations (\ref{55})-(\ref{57}) presents
the necessary and sufficient conditions for smooth matching of
interior as well as exterior metrics at star's surface. In the case
of pressure like constraint solution, the anisotropic solution,
i.e., the expressions of $\tilde{\rho}^{eff}$, $\tilde{p}_{r}^{eff}$
and $\tilde{p}_{t}^{eff}$ are evaluated in the following forms
\begin{eqnarray}\nonumber
\tilde{\rho}^{eff}&=&\frac{1}{2r^{2}}+\frac{e^{-\mathcal{A}r^{2}}}{2}
\left(5\mathcal{A}-\frac{1}{r^{2}}-\mathcal{B}^{2}r^{2}+\mathcal{AB}
r^{2}\right)+\chi\left\{e^{-\mathcal{A}r^{2}}\left(\frac{1}{r^{2}}
-2\mathcal{A}\right)\right.\\\nonumber&+&\left.\frac{2\mathcal{B}r^{2}
-1}{r^{2}(1+2\mathcal{B}r^{2})^{2}}\right\}+\frac{6\sigma
e^{-2\mathcal{A}r^{2}}}{r^{4}}\left[-4+\mathcal{B}r^{2}(1-\mathcal{B}
r^{2}+\mathcal{B}^{2}r^{4})+e^{\mathcal{A}r^{2}}(4\right.\\\nonumber&+&
\left.\mathcal{A}r^{2}-\mathcal{B}r^{2})+6\mathcal{A}^{3}r^{6}(2
+\mathcal{B}r^{2})-\mathcal{A}^{2}r^{4}(16+33\mathcal{B}r^{2}
+6\mathcal{B}^{2}r^{4})+\mathcal{A}r^{2}\right.\\\label{58}&\times&\left.
(-5+17\mathcal{B}r^{2}+14\mathcal{B}^{2}r^{4})\right],\\\nonumber
\tilde{p}_{r}^{eff}&=&\frac{-1}{2r^{2}}+\frac{e^{-\mathcal{A}r^{2}}}{2}
\left(4\mathcal{B}+\frac{1}{r^{2}}-\mathcal{A}-\mathcal{AB}r^{2}
+\mathcal{B}^{2}r^{2}\right)-\frac{2\sigma e^{-2\mathcal{A}r^{2}}}
{r^{2}}\left[6\mathcal{A}^{3}r^{4}(2\right.\\\nonumber&+&\left.
\mathcal{B}r^{2})-\mathcal{A}^{2}r^{2}(20+43\mathcal{B}r^{2}
+10\mathcal{B}^{2}r^{4})+\mathcal{A}(4\mathcal{B}^{3}r^{6}
+34\mathcal{B}^{2}r^{4}+35\mathcal{B}r^{2}\right.\\\nonumber&-&\left.3
+3e^{\mathcal{A}r^{2}})-\mathcal{B}(-3+11\mathcal{B}r^{2}
+5\mathcal{B}^{2}r^{4}+3e^{\mathcal{A}r^{2}})\right]+\chi\left[\frac{1}
{r^{2}}-e^{-\mathcal{A}r^{2}}(2\mathcal{B}\right.\\\nonumber&+&\left.
\frac{1}{r^{2}})+\frac{2\sigma e^{-2\mathcal{A} r^{2}}}{r^{2}}
\left\{\frac{7}{r^{2}}-3\mathcal{A}^{2}r^{2}(4+4\mathcal{B}
r^{2}-\mathcal{B}^{2}r^{4})+2\mathcal{A}(16\mathcal{B}r^{2}+9
\mathcal{B}^{2}\right.\right.\\\nonumber&\times&\left.\left.r^{4}
+\mathcal{B}^{3}r^{6}+4)-\mathcal{B}(11\mathcal{B}r^{2}
+2\mathcal{B}^{2}r^{4}-\mathcal{B}^{3}r^{6})-\frac{e^{2\mathcal{A}
r^{2}}}{r^{2}}-\frac{6e^{\mathcal{A}r^{2}}}{r^{2}}\right\}\right],
\\\label{59}\\\nonumber
\tilde{p}_{t}^{eff}&=&-\frac{1}{2r^{2}}+\frac{e^{-\mathcal{A}r^{2}}}{2}
\left(\frac{1}{r^{2}}+\mathcal{B}^{2}r^{2}-\mathcal{AB}r^{2}
-\mathcal{A}+4\mathcal{B}\right)+\chi\left\{e^{-\mathcal{A} r^{2}}
(\mathcal{A}-2\mathcal{B}\right.\\\nonumber&+&\left.\mathcal{AB}r^{2}
-\mathcal{B}^{2}r^{2})+\frac{\mathcal{B}^{2}r^{2}(2\mathcal{B}
r^{2}-3)}{(1+2\mathcal{B}r^{2})^{2}}\right\}-\frac{2\sigma
e^{-2\mathcal{A}r^{2}}}{r^{2}}\left[\mathcal{A}(-3+3e^{\mathcal{A}r^{2}}
+35\right.\\\nonumber&\times&\left.\mathcal{B}r^{2}+34\mathcal{B}^{2}
r^{4}+4\mathcal{B}^{3}r^{6})+6\mathcal{A}^{3}r^{4}(2+ \mathcal{B}
r^{2})-\mathcal{A}^{2}r^{2}(20+43\mathcal{B}r^{2} +10\right.
\\\label{60}&\times&\left.\mathcal{B}^{2}r^{4})-\mathcal{B}(-3
+11\mathcal{B}r^{2}+5\mathcal{B}^{2}r^{4}+3e^{\mathcal{A}r^{2}})\right].
\end{eqnarray}
The anisotropic factor in this case is calculated as
\begin{eqnarray}\nonumber
\tilde{\Delta}^{eff}&=&\chi\left[\frac{2\mathcal{B}^{3}r^{6}
-\mathcal{B}^{2}r^{4}-4\mathcal{B}r^{2}-1}{r^{2}(1+2
\mathcal{B}r^{2})^{2}}+e^{-\mathcal{A}r^{2}}\left(\mathcal{A}
+\frac{1}{r^{2}}-\mathcal{B}^{2}r^{2}+\mathcal{A}\mathcal{B}
r^{2}\right)\right.\\\nonumber&+&\left.\frac{2\sigma
e^{-2\mathcal{A}r^{2}}}{r^{2}}\left\{\frac{e^{2\mathcal{A}
r^{2}}}{r^{2}}+\frac{6e^{\mathcal{A}r^{2}}}{r^{2}}-\frac{7}{r^{2}}
+3\mathcal{A}^{2}r^{2}(4+4\mathcal{B}r^{2}+\mathcal{B}^{2}r^{4})
-2\mathcal{A}\right.\right.\\\label{61}&\times&\left.\left.(4
+16\mathcal{B}r^{2}+9\mathcal{B}^{2}r^{4}+\mathcal{B}^{3}r^{6})
+\mathcal{B}^{2}r^{2}(11+2\mathcal{B}r^{2}-\mathcal{B}^{2}
r^{4})\right\}\right].
\end{eqnarray}
\begin{figure}\center
\epsfig{file=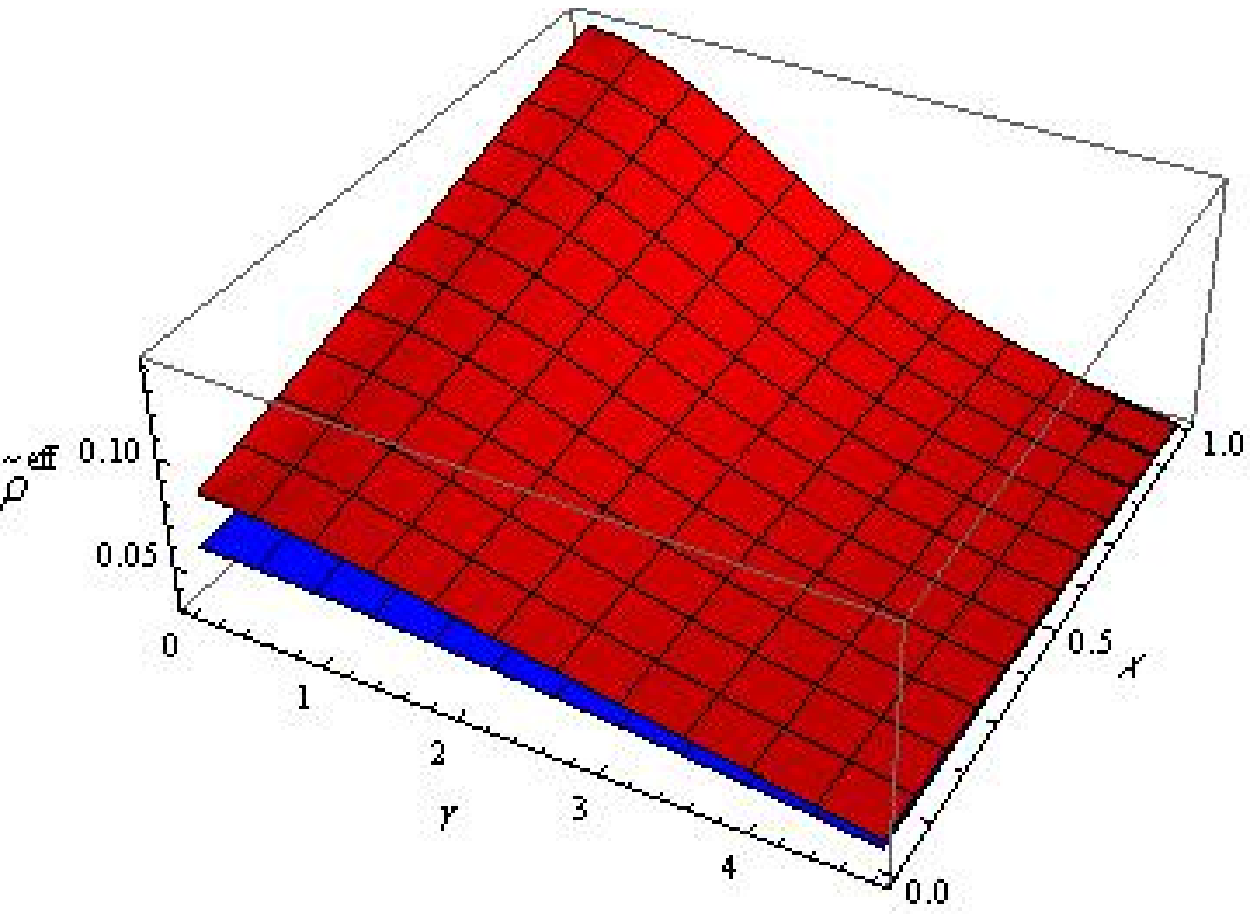,width=0.45\linewidth}
\epsfig{file=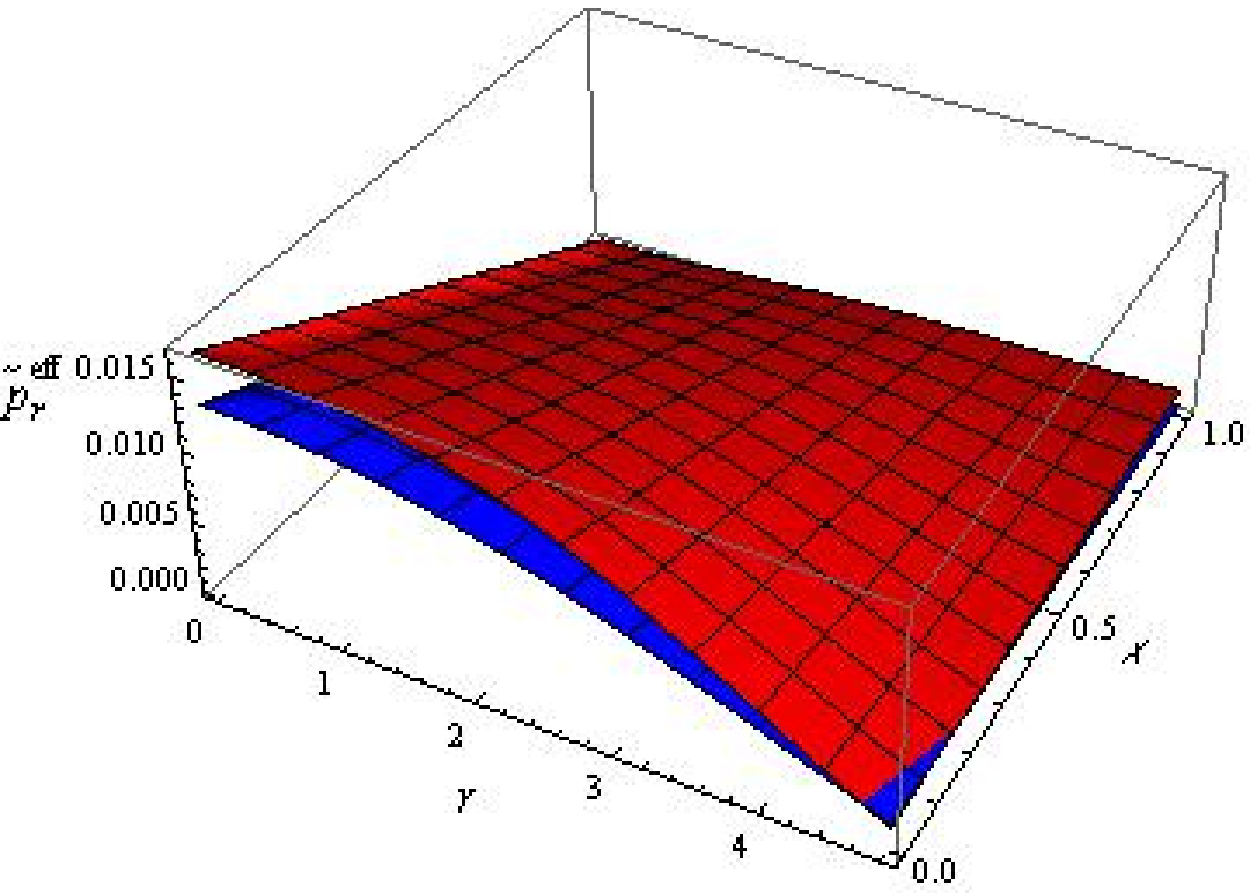,width=0.45\linewidth}
\epsfig{file=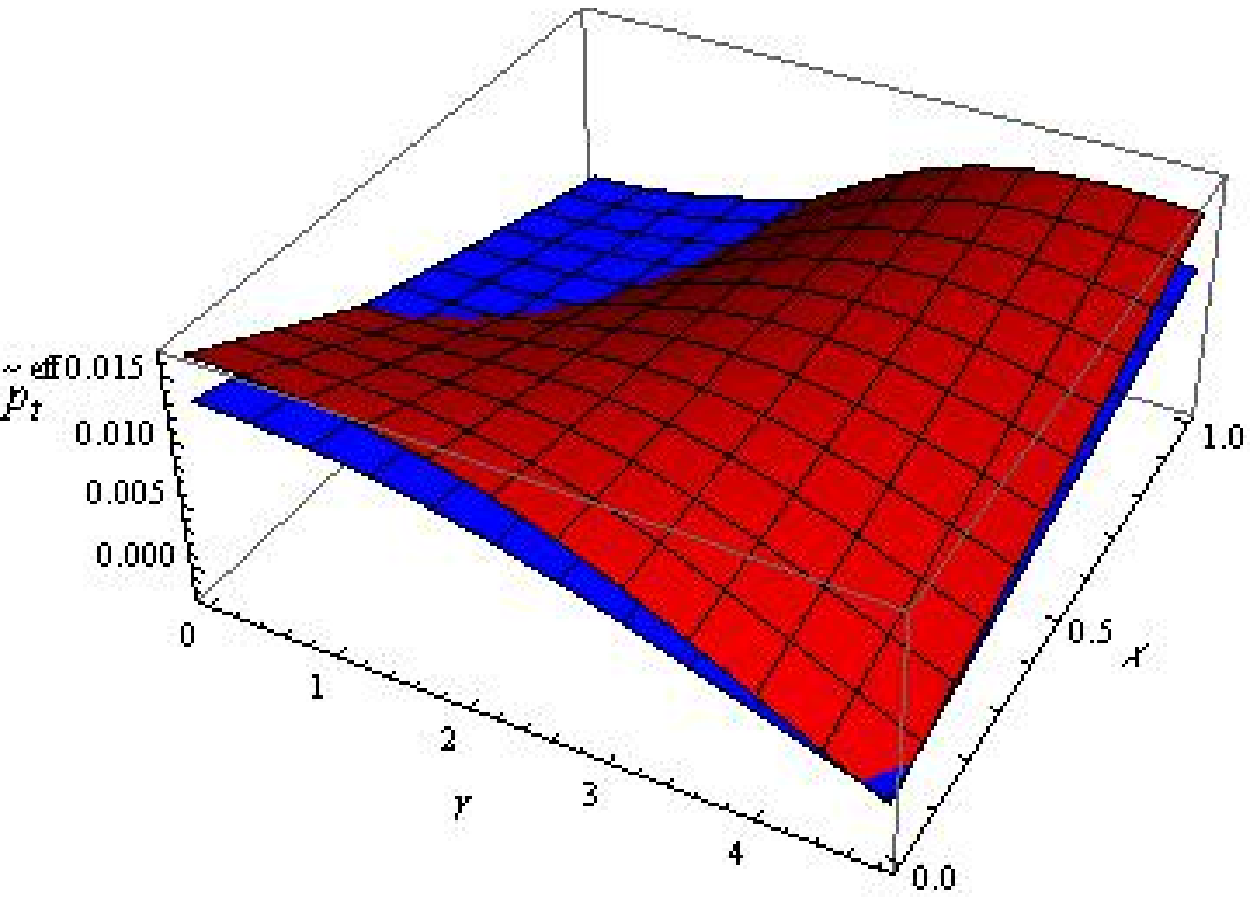,width=0.45\linewidth}
\epsfig{file=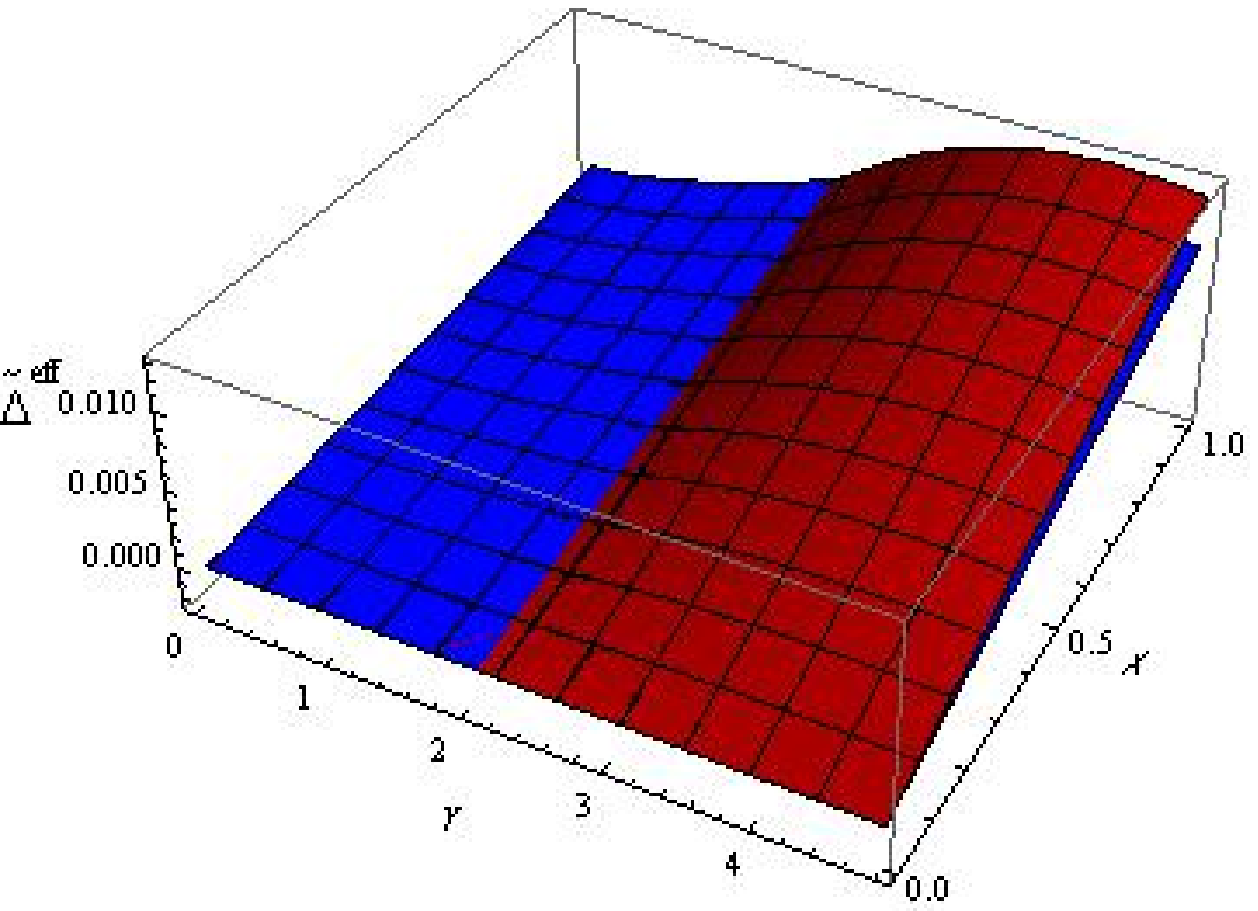,width=0.45\linewidth}\\
\caption{Plots of $\tilde{\rho}^{eff}$, $\tilde{p}_{r}^{eff}$,
$\tilde{p}_{t}^{eff}$ and $\tilde{\Delta}^{eff}$ versus $r$ and
$\chi$ with $\sigma=2$, $Q_{0}=0.1$ (red), $Q_{0}=1$ (blue),
$M_{0}=1M_{\odot}$ and $\mathcal{R}=4M_{\odot}$ for the first
solution.}
\end{figure}

In order to examine the physical properties of stellar system
corresponding to the first solution, we analyze their graphical
behavior for $\sigma=2$ and two different values of $Q_{0}$. In this
regard, we employ the constant $\mathcal{A}$ as presented in
Eq.(\ref{55}) whereas $\mathcal{B}$ is assumed as a free parameter
and will be considered from Eq.(\ref{48}). The structure of
self-gravitating objects demands that the nature of energy density
as well as radial pressure should be finite, positive, maximum and
regular in the interior of compact objects. The physical analysis of
effective energy density and effective pressure (radial and
tangential) is presented in Figure \textbf{1}. The effective energy
density shows maximum behavior at the center of star and decreases
monotonically with increase in $r$. We notice that the larger value
of $Q_{0}$ yields smaller $\tilde{\rho}^{eff}$ which indicates that
increase in charge makes the sphere less dense. It is also found
that the value of $\tilde{\rho}^{eff}$ enhances with increasing
$\chi$.

The behavior of $\tilde{p}_{r}^{eff}$ as well as
$\tilde{p}_{t}^{eff}$ in the presence of charge is also regular as
well as finite similar to that of effective energy density. The
value of $\tilde{p}_{r}^{eff}$ becomes zero at the boundary of
star's surface and represents decreasing behavior with increase in
$Q_{0}$, $r$ and $\chi$. The physical interpretation of
$\tilde{p}_{t}^{eff}$ also reveals monotonically decreasing behavior
with respect to $r$. The role of effective anisotropic factor with
the inclusion of charge is also examined graphically in Figure
\textbf{1} which shows that the variation of $\tilde{\Delta}^{eff}$
remains positive. This indicates the existence of a repellent source
that permits the evolution of large massive configuration in the
interior region of stellar object. This factor depicts a constant
behavior for small values of $\chi$ but with the increase in $\chi$,
the value of $\tilde{\Delta}^{eff}$ decreases for larger value of
$Q_{0}$.

To check physical consistency of the resulting solutions and
existence of ordinary matter configuration, there are some physical
features known as energy conditions. These conditions are the
constraints imposed on the energy-momentum tensor and are
categorized into null, strong, weak and dominant energy conditions.
In $f(R)$ gravity with the influence of charge, these conditions for
anisotropic matter distribution are expressed as
\begin{itemize}
\item NEC: \quad$\tilde{\rho}^{eff}+\tilde{p}_{r}^{eff}\geq 0$,\quad
$\tilde{\rho}^{eff}+\tilde{p}_{t}^{eff}+\frac{q^{2}}{4\pi r^{4}}\geq
0$,
\item SEC: \quad$\tilde{\rho}^{eff}+\tilde{p}_{r}^{eff}+
2\tilde{p}_{t}^{eff}+\frac{q^{2}}{4\pi r^{4}}\geq 0$,
\item WEC:\quad$\tilde{\rho}^{eff}+\frac{q^{2}}{8\pi r^{4}}\geq 0$,
\quad$\tilde{\rho}^{eff}+\tilde{p}_{r}^{eff}\geq
0$,\quad$\tilde{\rho}^{eff}+ \tilde{p}_{t}^{eff}+\frac{q^{2}}{4\pi
r^{4}}\geq 0$,
\item DEC:
\quad$\tilde{\rho}^{eff}-\tilde{p}_{r}^{eff}+\frac{q^{2}}{4\pi
r^{4}}\geq 0$,\quad$\tilde{\rho}^{eff}- \tilde{p}_{t}^{eff}\geq0$.
\end{itemize}
Figure \textbf{2} represents that all energy conditions are
satisfied which assure the physical viability of the considered
charged anisotropic solution.
\begin{figure}\center
\epsfig{file=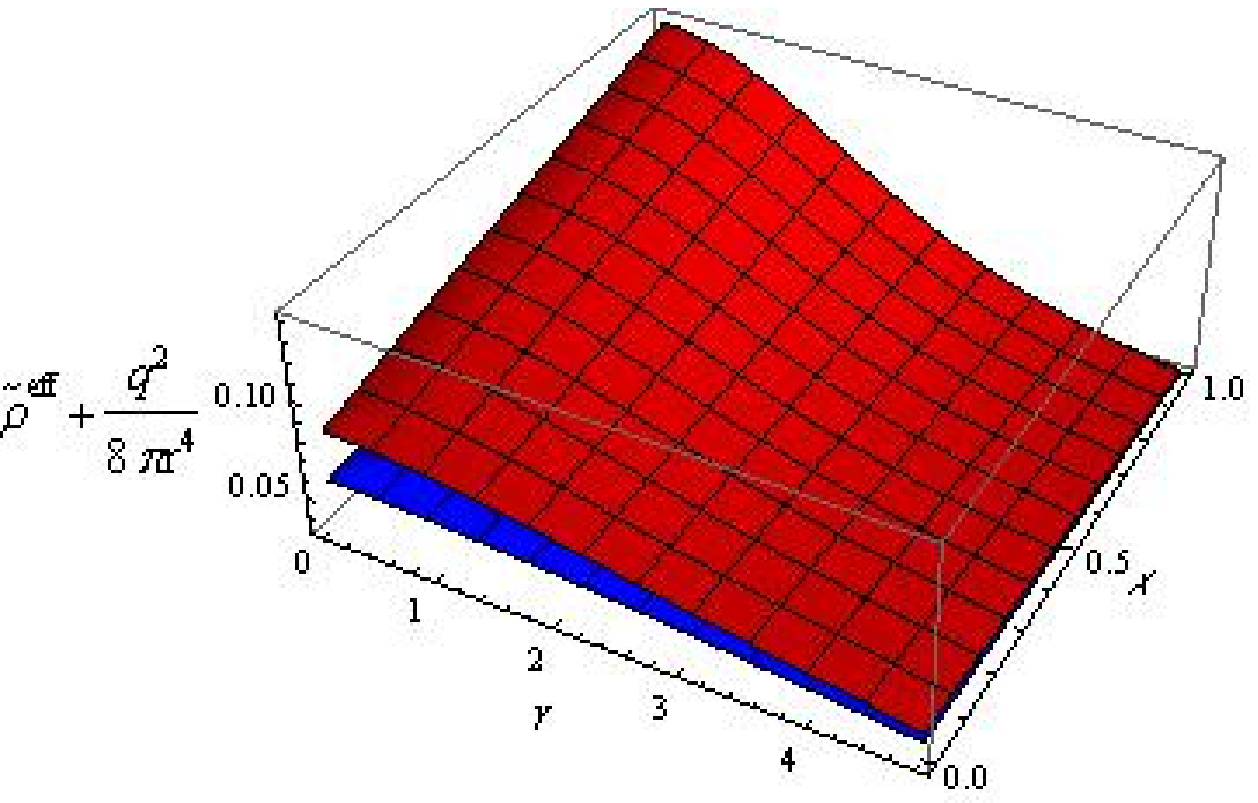,width=0.48\linewidth}
\epsfig{file=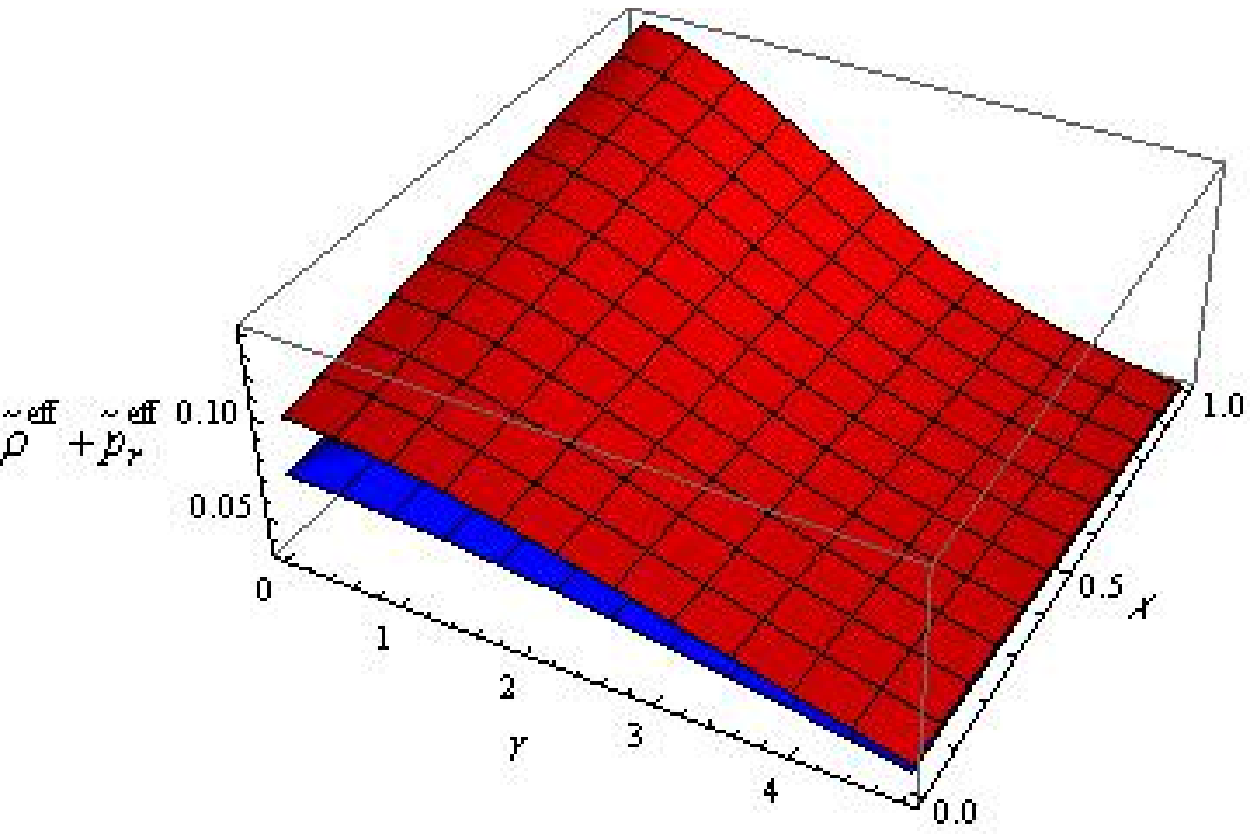,width=0.45\linewidth}
\epsfig{file=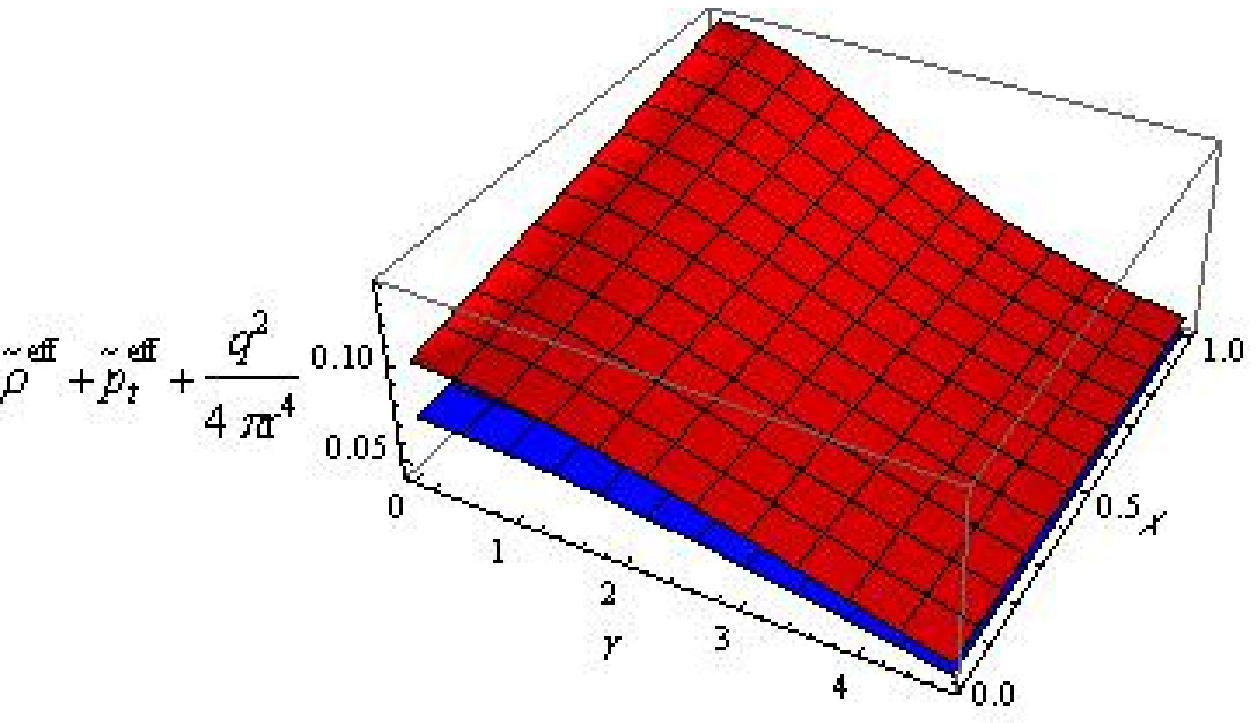,width=0.48\linewidth}
\epsfig{file=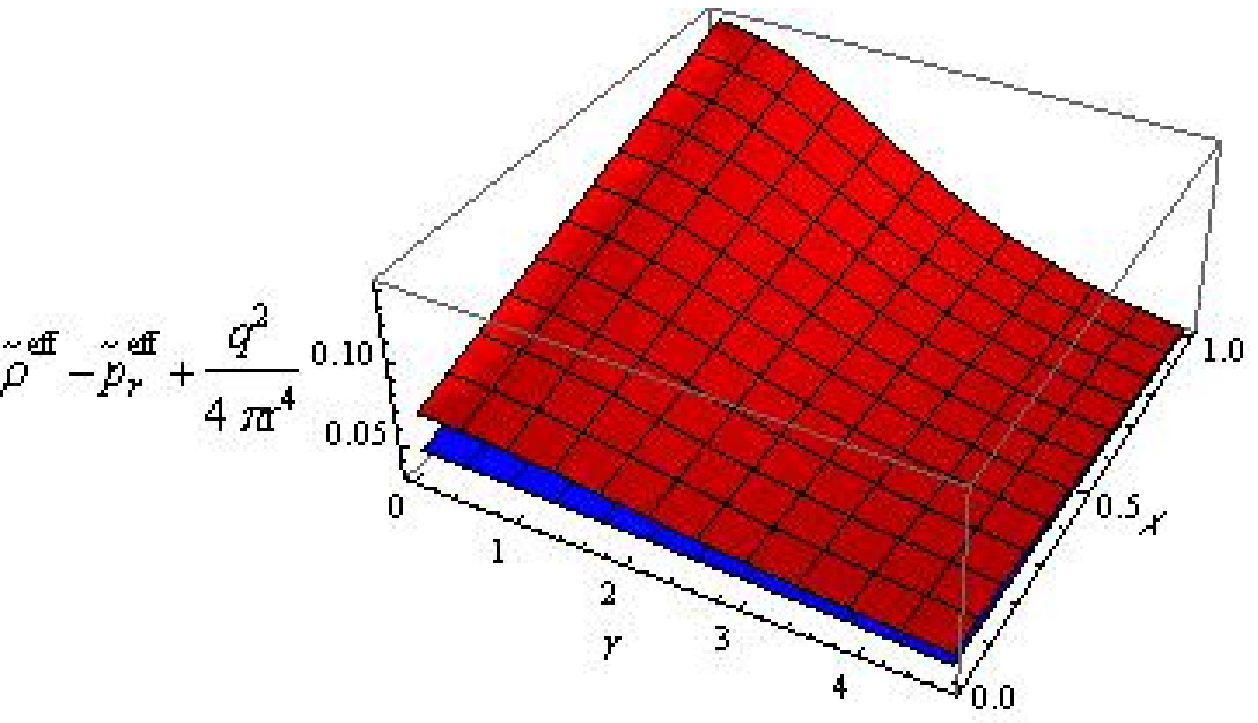,width=0.48\linewidth}
\epsfig{file=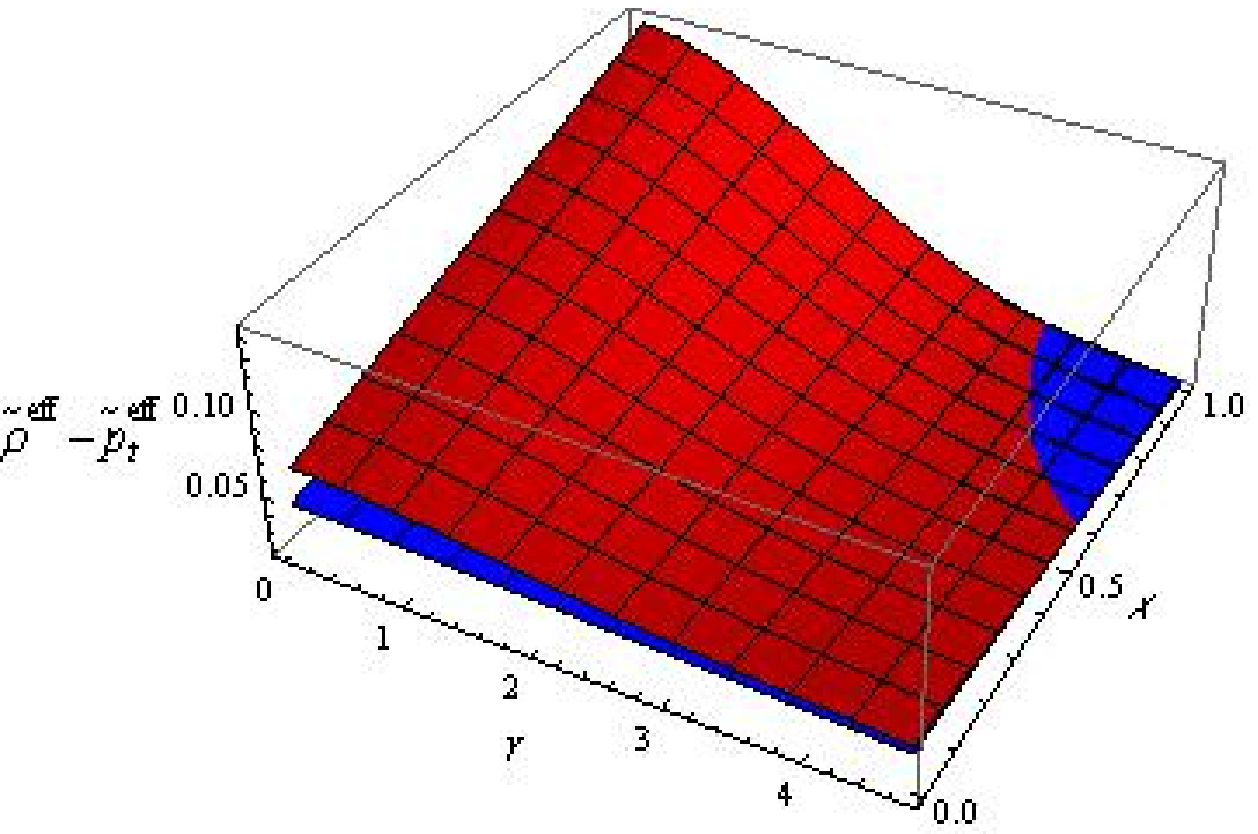,width=0.44\linewidth}
\epsfig{file=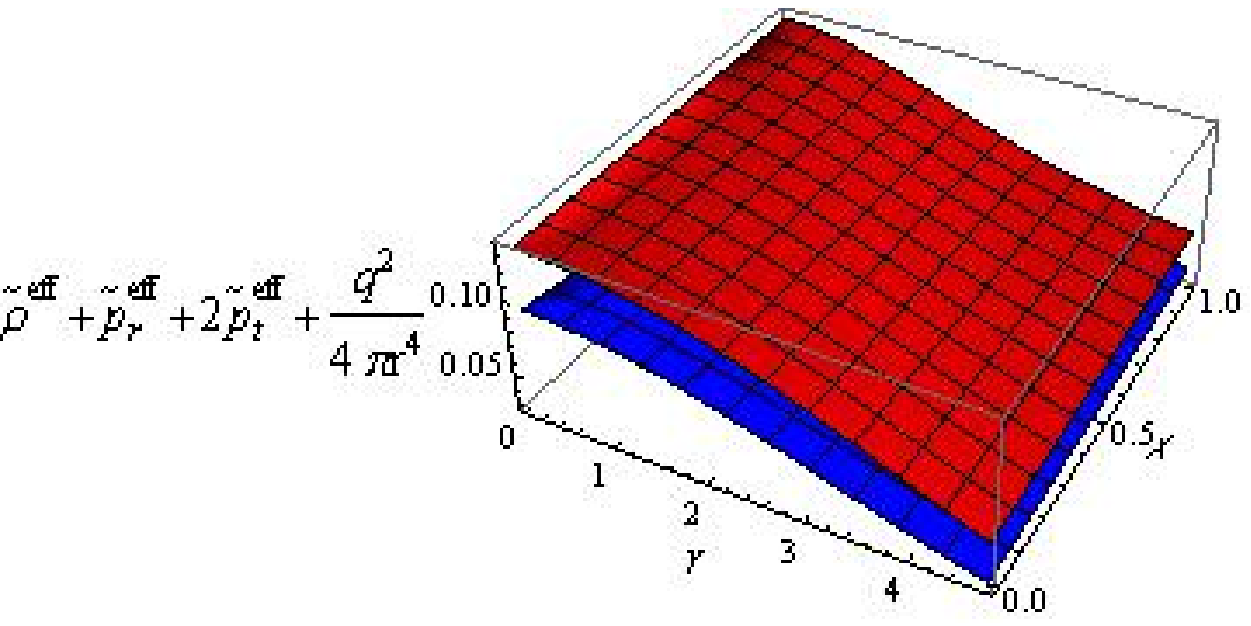,width=0.51\linewidth}\\
\caption{Behavior of energy conditions versus $r$ and $\chi$ with
$\sigma=2$, $Q_{0}=0.1$ (red), $Q_{0}=1$ (blue), $M_{0}=1M_{\odot}$
and $\mathcal{R}=4M_{\odot}$ for the first solution.}
\end{figure}
\begin{figure}\center
\epsfig{file=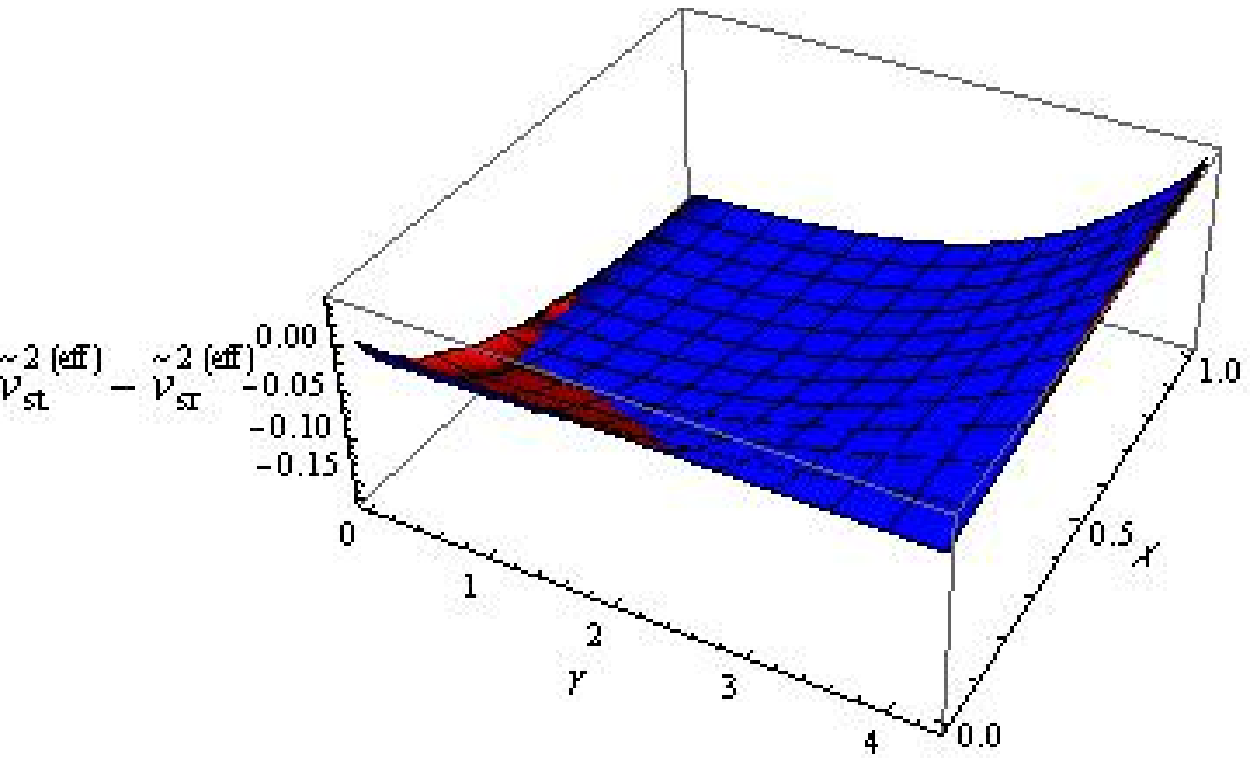,width=0.5\linewidth}\\
\caption{Plots of $\tilde{v}^{2(eff)}_{st}-\tilde{v}^{2(eff)}_{sr}$
versus $r$ and $\chi$ with $\sigma=2$, $Q_{0}=0.1$ (red), $Q_{0}=1$
(blue), $M_{0}=1M_{\odot}$ and $\mathcal{R}=4M_{\odot}$ for the
first solution.}
\end{figure}

In astrophysics, the stability of stellar structure has a crucial
role in evaluating any physically viable system. We discuss the
stability of charged anisotropic solution by means of squared speed
of sound ($v^{2}_{s}$) based on Herrera's cracking concept
\cite{40}. The causality condition demands that the squared speed of
sound represented by $v^{2}_{s} = dp/d\rho$ must be in the range [0,
1], i.e., $0 \leq v^{2}_{s} \leq 1$ in the interior geometry of
stars for a physically stable structure. Herrera \cite{40} described
the idea of cracking by considering a different technique to obtain
potentially stable or unstable regions of compact stars. These
regions are evaluated by the difference of squared sound speed in
tangential and radial directions as $|v^{2}_{st}-v^{2}_{sr}|\leq1$,
where $v^{2}_{sr}$ and $v^{2}_{st}$ indicate squared sound speed in
the transverse and radial directions, respectively. The stability
analysis for $Q_{0}=0.1,1$ is shown in Figure \textbf{3} which
interprets that our resulting charged anisotropic solution is stable
for all adopted values of $\sigma$, $Q_{0}$ and $\chi$. It is also
observed that there is a very small increase in stability with
increase in $Q_{0}$.

\subsection{The Second Solution}

Here, we consider an alternative form of constraint to attain a
second type of physically acceptable charged anisotropic solution.
We take a density like constraint \cite{36} presented by
\begin{equation}\nonumber
\Theta_{0}^{0}=\rho^{eff}+\frac{q^{2}}{8\pi r^{4}}.
\end{equation}
From Eqs.(\ref{24}) and (\ref{28}), we evaluate
\begin{equation}\nonumber
I^{*'}+\frac{I^{*}}{r}+\frac{1}{r}+{e^{-\mathcal{A}r^{2}}}
\left(2\mathcal{A}r-\frac{1}{r}\right)=0,
\end{equation}
whose solution is obtained as
\begin{equation}\nonumber
I^{*}=\frac{d_{1}}{r}-1+e^{-\mathcal{A}r^{2}},
\end{equation}
where $d_{1}$ shows the integration constant. In order to have a
singularity free solution at the center ($r=0$) of stellar object,
we suppose that $d_{1}=0$ which leads to
\begin{equation}\label{62}
I^{*}=e^{-\mathcal{A}r^{2}}-1.
\end{equation}
By adopting the same strategy as applied for the first solution, the
matching conditions are given as
\begin{eqnarray}\label{63}
\frac{2M}{\mathcal{R}}&=&\frac{2M_{0}}{\mathcal{R}}
+\frac{\mathcal{Q}^{2}-Q_{0}^{2}}{\mathcal{R}^{2}}+
\chi(1-e^{-\mathcal{A}r^{2}}),\\\label{64}
\mathcal{B}\mathcal{R}^{2}+\mathcal{C}&=&\ln\left[1
-\frac{2M_{0}}{\mathcal{R}}+\frac{Q_{0}^{2}}{R^{2}}
-\chi(1-e^{-\mathcal{A}r^{2}})\right].
\end{eqnarray}
The expressions of charged anisotropic solution in terms of
$\tilde{\rho}^{eff}$, $\tilde{p}_{r}^{eff}$ and
$\tilde{p}_{t}^{eff}$ for density like constraint are obtained as
follows
\begin{eqnarray}\nonumber
\tilde{\rho}^{eff}&=&\frac{1}{2r^{2}}+\frac{e^{-\mathcal{A}r^{2}}}{2}
\left(5\mathcal{A}-\frac{1}{r^{2}}-\mathcal{B}^{2}r^{2}+\mathcal{AB}
r^{2}\right)+\frac{6\sigma e^{-2\mathcal{A}r^{2}}}{r^{4}}
\left\{-4+\mathcal{B}r^{2}(1\right.\\\nonumber&-&\left.\mathcal{B}
r^{2}+\mathcal{B}^{2}r^{4})+e^{\mathcal{A}r^{2}}(4+\mathcal{A}r^{2}
-\mathcal{B}r^{2})+6\mathcal{A}^{3}r^{6}(2+\mathcal{B}r^{2})
-\mathcal{A}^{2}r^{4}(33\mathcal{B}r^{2}\right.\\\nonumber&+&\left.6
\mathcal{B}^{2}r^{4}+16)+\mathcal{A}r^{2}(-5+17\mathcal{B} r^{2}
+14\mathcal{B}^{2}r^{4})\right\}+\chi\left[e^{-\mathcal{A}r^{2}}
\left(2\mathcal{A}-\frac{1}{r^{2}}\right)\right.\\\nonumber&+&\left.
\frac{1}{r^{2}}+\frac{2\sigma e^{-2\mathcal{A}r^{2}}}{r^{4}}
\left\{2(1-e^{2\mathcal{A}r^{2}}) +\mathcal{B}r^{2}(-3-3\mathcal{B}
r^{2}+9\mathcal{B}^{2}r^{4}+2\mathcal{B}^{3}r^{6})\right.\right.
\\\nonumber&+&\left.\left.3e^{\mathcal{A}r^{2}}(\mathcal{B}r^{2}
-\mathcal{A}r^{2})+6\mathcal{A}^{3}r^{6}(2 +\mathcal{B}r^{2})
-\mathcal{A}^{2}r^{4}(37\mathcal{B}r^{2}+4\mathcal{B}^{2}r^{4}+32)
\right.\right.\\\label{65}&+&\left.\left.\mathcal{A}r^{2}
(7+45\mathcal{B}r^{2}+10\mathcal{B}^{2}r^{4}-4\mathcal{B}^{3}
r^{6})\right\}\right],\\\nonumber
\tilde{p}_{r}^{eff}&=&\frac{-1}{2r^{2}}+\frac{e^{-\mathcal{A}r^{2}}}{2}
\left(4\mathcal{B}+\frac{1}{r^{2}}-\mathcal{A}-\mathcal{AB}r^{2}
+\mathcal{B}^{2}r^{2}\right)+\frac{\chi}{r^{2}}(1+2\mathcal{B}r^{2})
(e^{-\mathcal{A}r^{2}}\\\nonumber&-&1)-\frac{2\sigma
e^{-2\mathcal{A}r^{2}}} {r^{2}}\left[6\mathcal{A}^{3}r^{4}(2+
\mathcal{B}r^{2})-\mathcal{A}^{2}r^{2}(20+43\mathcal{B}r^{2}
+10\mathcal{B}^{2}r^{4})+\mathcal{A}\right.\\\nonumber&\times&\left.(4
\mathcal{B}^{3}r^{6}+34\mathcal{B}^{2}r^{4}+35\mathcal{B}r^{2}-3
+3e^{\mathcal{A}r^{2}})-\mathcal{B}(-3+11\mathcal{B}r^{2}+5
\mathcal{B}^{2}r^{4}\right.\\\label{66}&+&\left.3
e^{\mathcal{A}r^{2}})\right],\\\nonumber
\tilde{p}_{t}^{eff}&=&-\frac{1}{2r^{2}}+\frac{e^{-\mathcal{A}r^{2}}}{2}
\left(\frac{1}{r^{2}}+\mathcal{B}^{2}r^{2}-\mathcal{AB}r^{2}
-\mathcal{A}+4\mathcal{B}\right)+\chi\left\{e^{-\mathcal{A} r^{2}}
(2\mathcal{B}-\mathcal{A}\right.\\\nonumber&-&\left.\mathcal{AB}r^{2}
+\mathcal{B}^{2}r^{2})-2\mathcal{B}-\mathcal{B}^{2}r^{2}\right\}
-\frac{2\sigma e^{-2\mathcal{A}r^{2}}}{r^{2}}\left[\mathcal{A}
(35\mathcal{B}r^{2} +34\mathcal{B}^{2}r^{4}+4\mathcal{B}^{3}
\right.\\\nonumber&\times&\left.r^{6}-3+3e^{\mathcal{A}r^{2}})
+6\mathcal{A}^{3}r^{4}(2+\mathcal{B}r^{2})-\mathcal{A}^{2}r^{2}(20
+43\mathcal{B}r^{2}+10\mathcal{B}^{2}r^{4})\right.\\\label{67}&-&\left.
\mathcal{B}(-3+11\mathcal{B}r^{2}+5\mathcal{B}^{2}r^{4}+3e^{\mathcal{A}
r^{2}})\right].
\end{eqnarray}
The anisotropic factor corresponding to the second solution becomes
\begin{equation}\label{68}
\tilde{\Delta}^{eff}=\chi\left\{e^{-\mathcal{A}r^{2}}
\left(\mathcal{B}^{2}r^{2}-\mathcal{A}-\mathcal{A}\mathcal{B}
r^{2}-\frac{1}{r^{2}}\right)+\frac{1-\mathcal{B}^{2}
r^{4}}{r^{2}}\right\}.
\end{equation}
This factor vanishes for $\chi=0$ and our solution reduces to the
standard isotropic solution.
\begin{figure}\center \epsfig{file=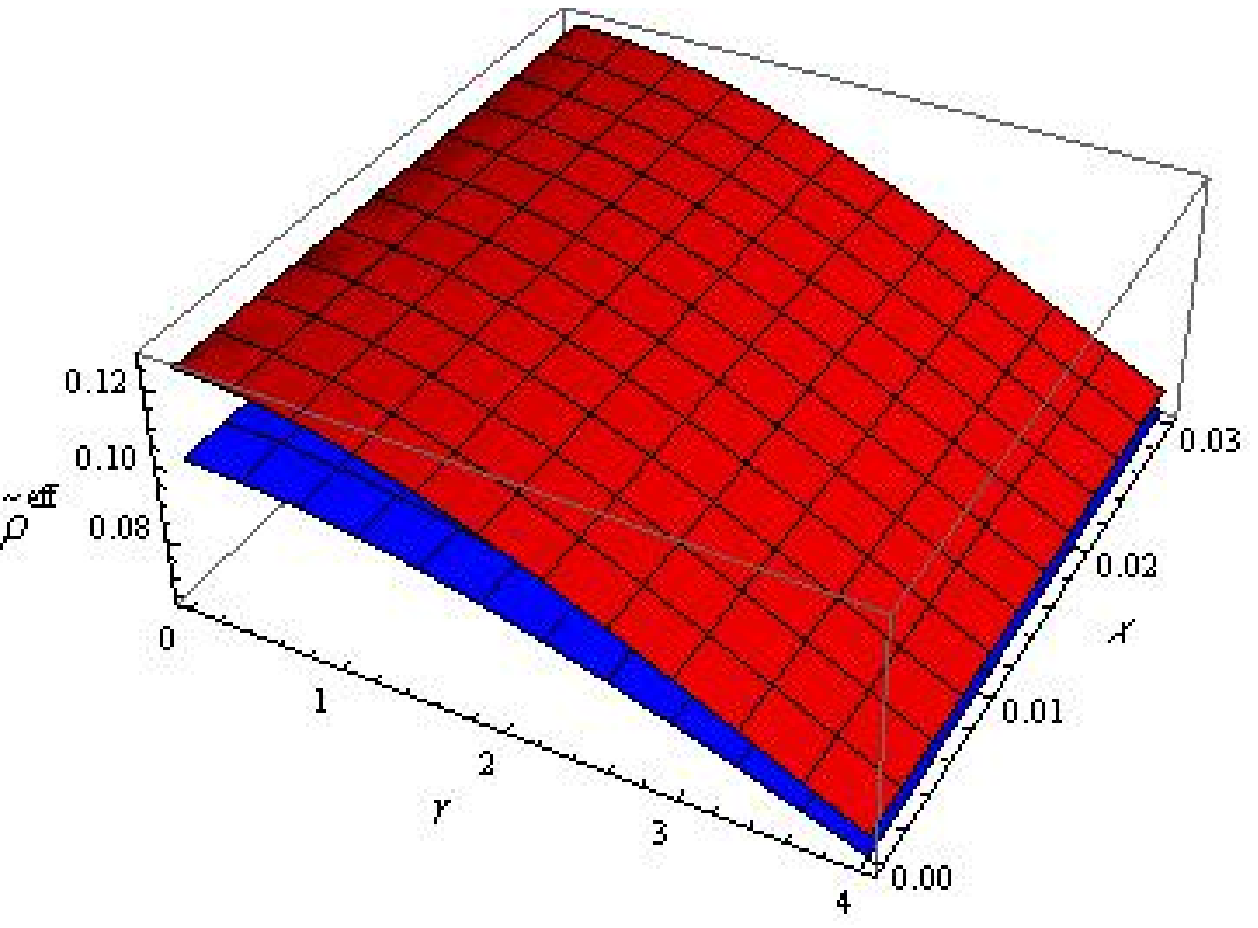,width=0.48\linewidth}
\epsfig{file=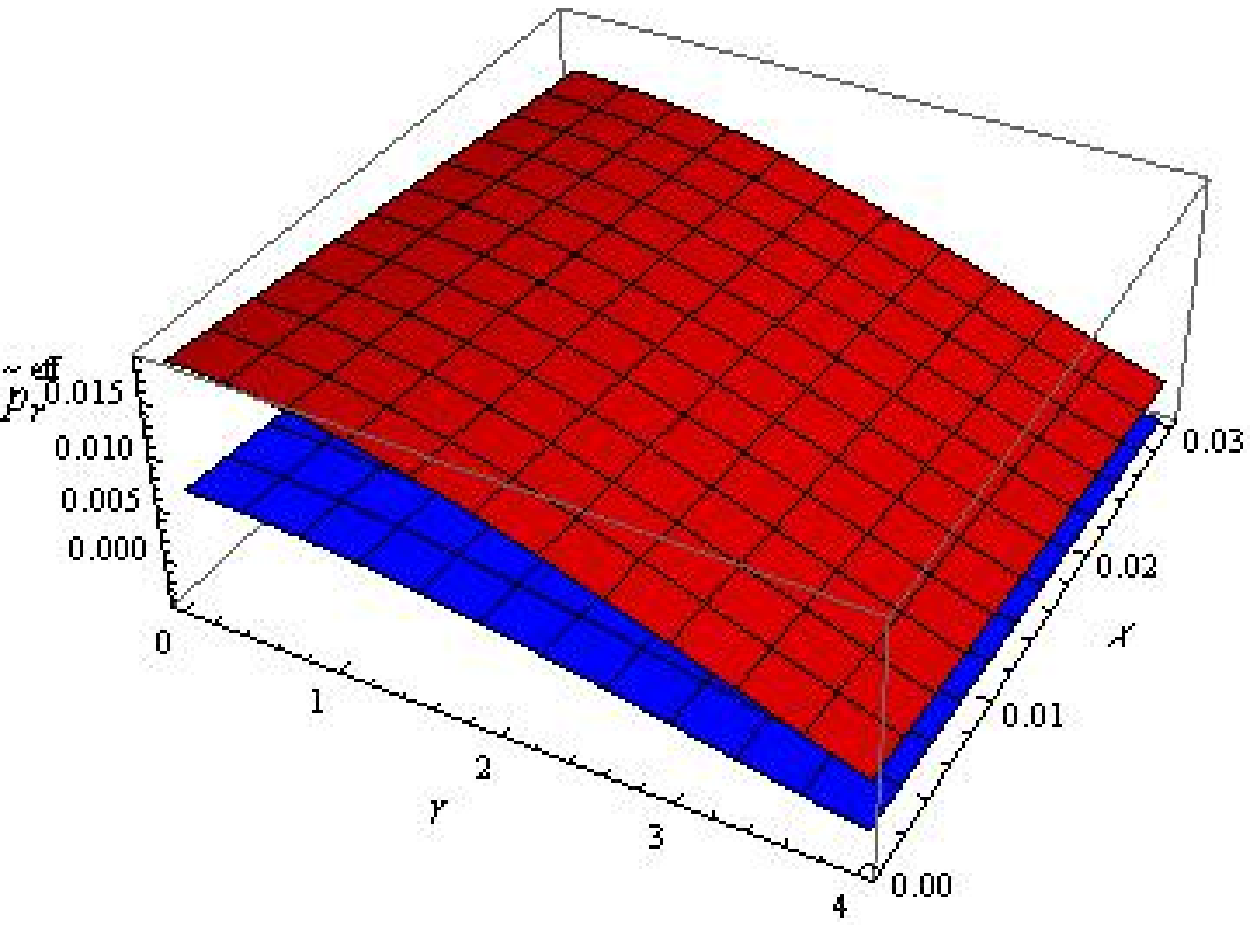,width=0.48\linewidth}
\epsfig{file=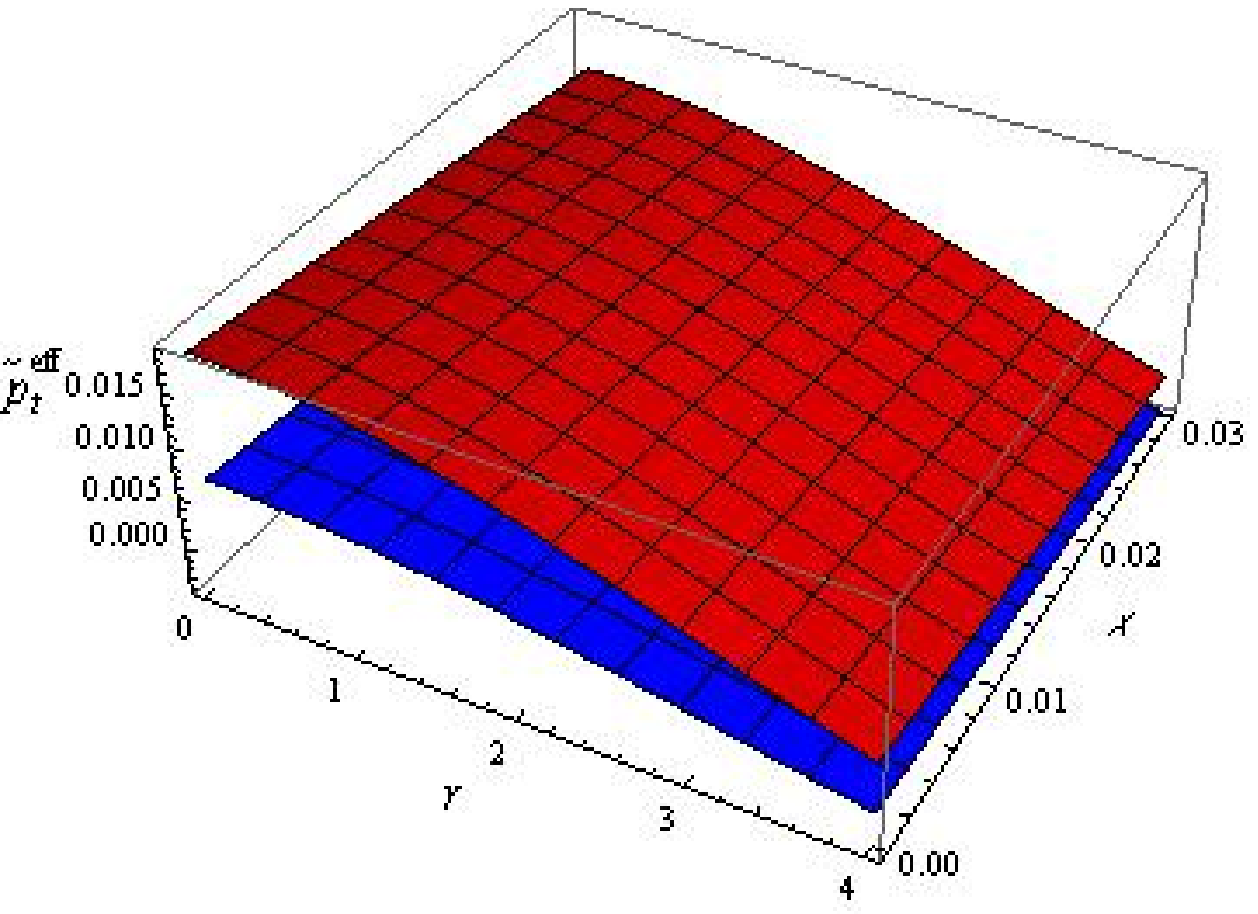,width=0.48\linewidth}
\epsfig{file=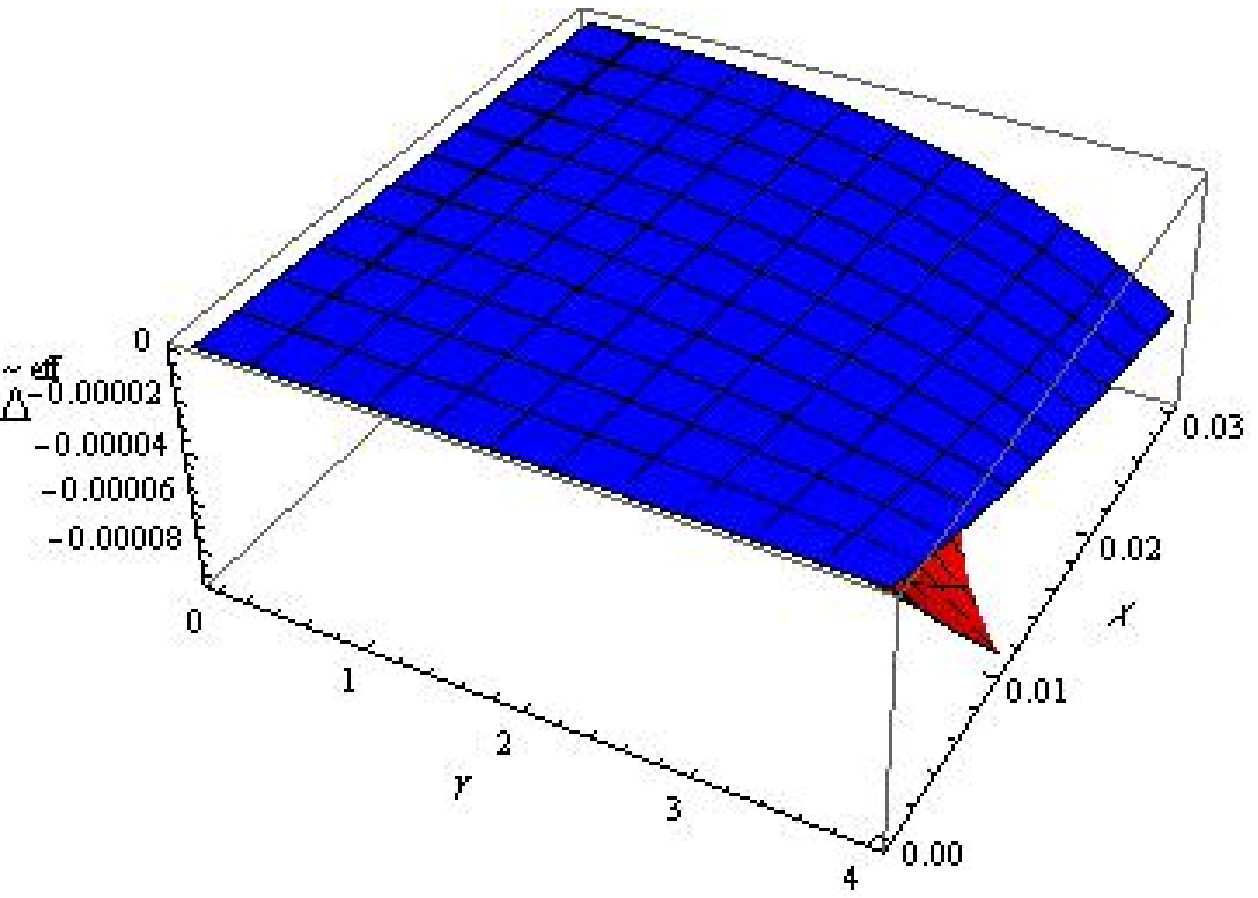,width=0.48\linewidth}\\
\caption{Plots of $\tilde{\rho}^{eff}$, $\tilde{p}_{r}^{eff}$,
$\tilde{p}_{t}^{eff}$ and $\tilde{\Delta}^{eff}$ versus $r$ and
$\chi$ with $\sigma=0.2$, $Q_{0}=0.1$ (red), $Q_{0}=1$ (blue),
$M_{0}=1M_{\odot}$ and $\mathcal{R}=4M_{\odot}$ for the second
solution.}
\end{figure}
\begin{figure}\center
\epsfig{file=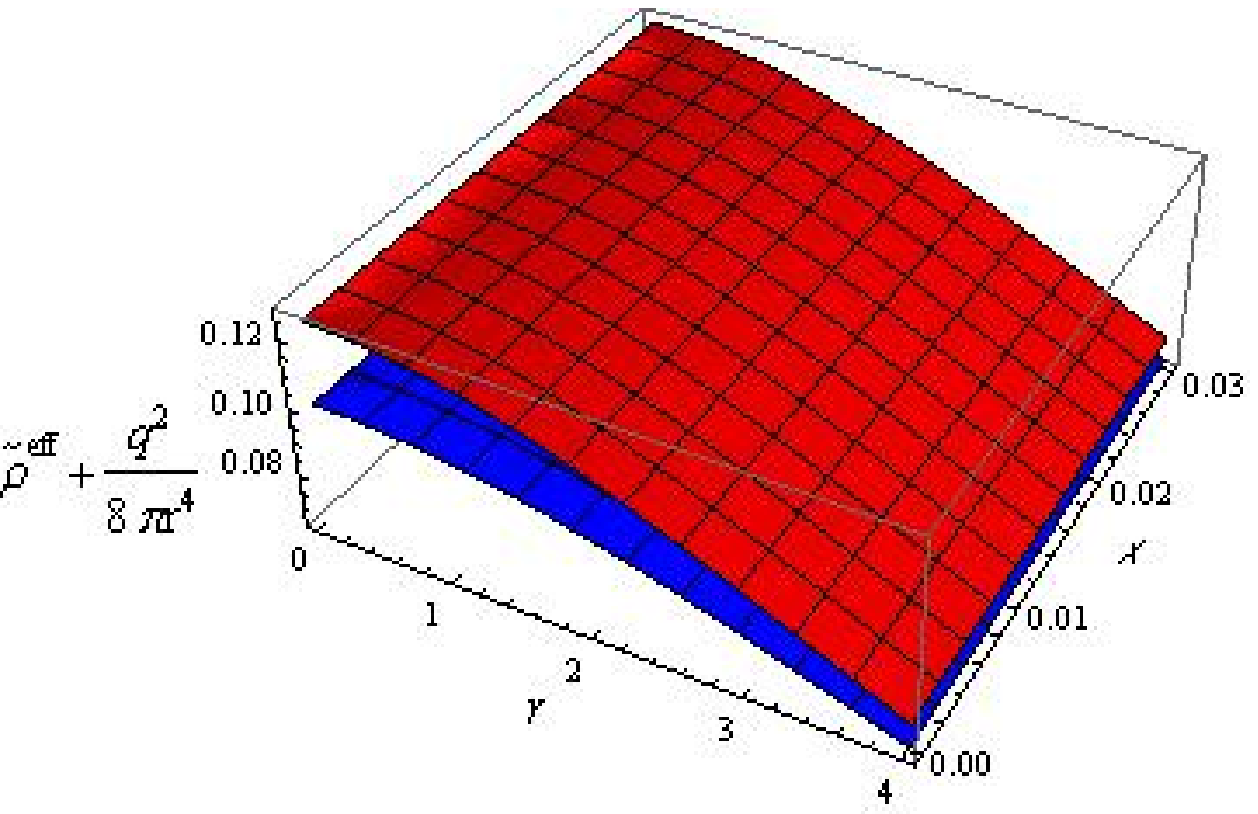,width=0.48\linewidth}
\epsfig{file=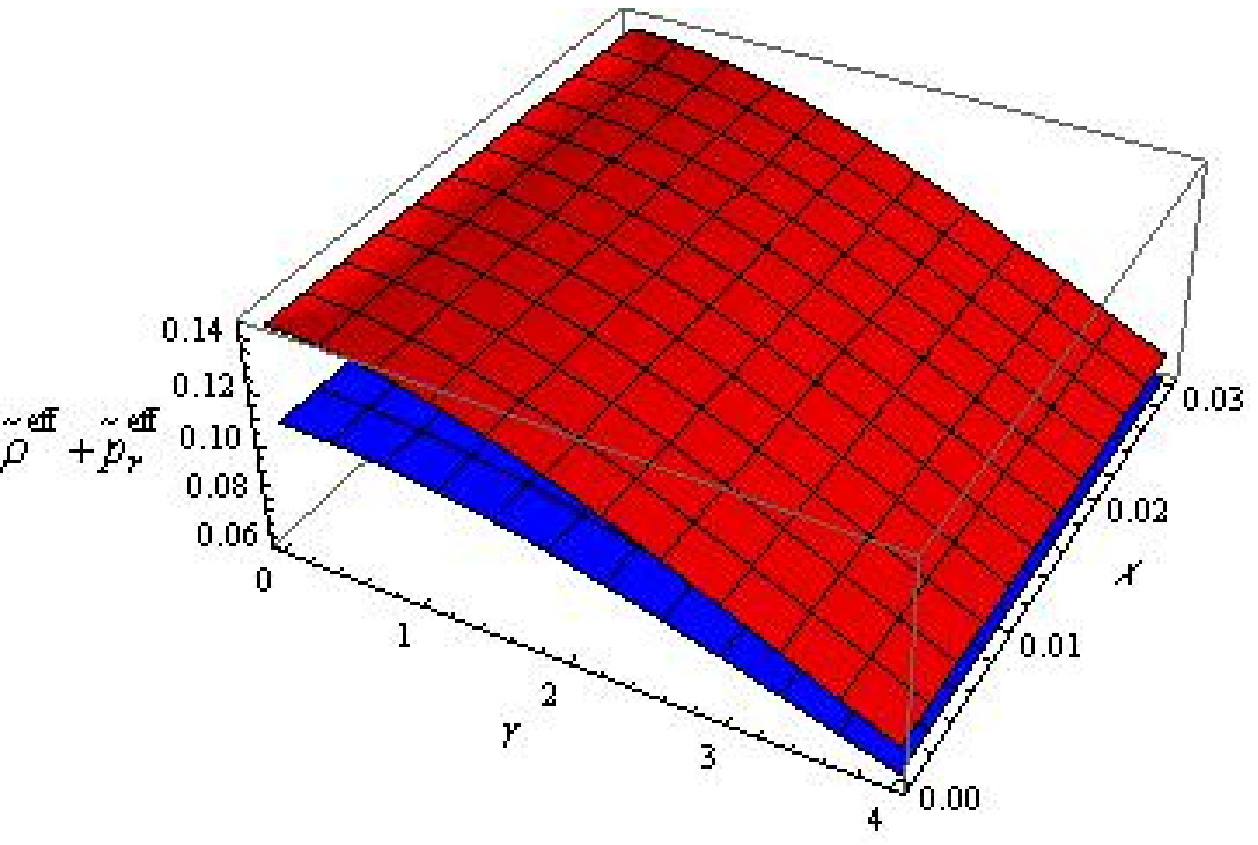,width=0.46\linewidth}
\epsfig{file=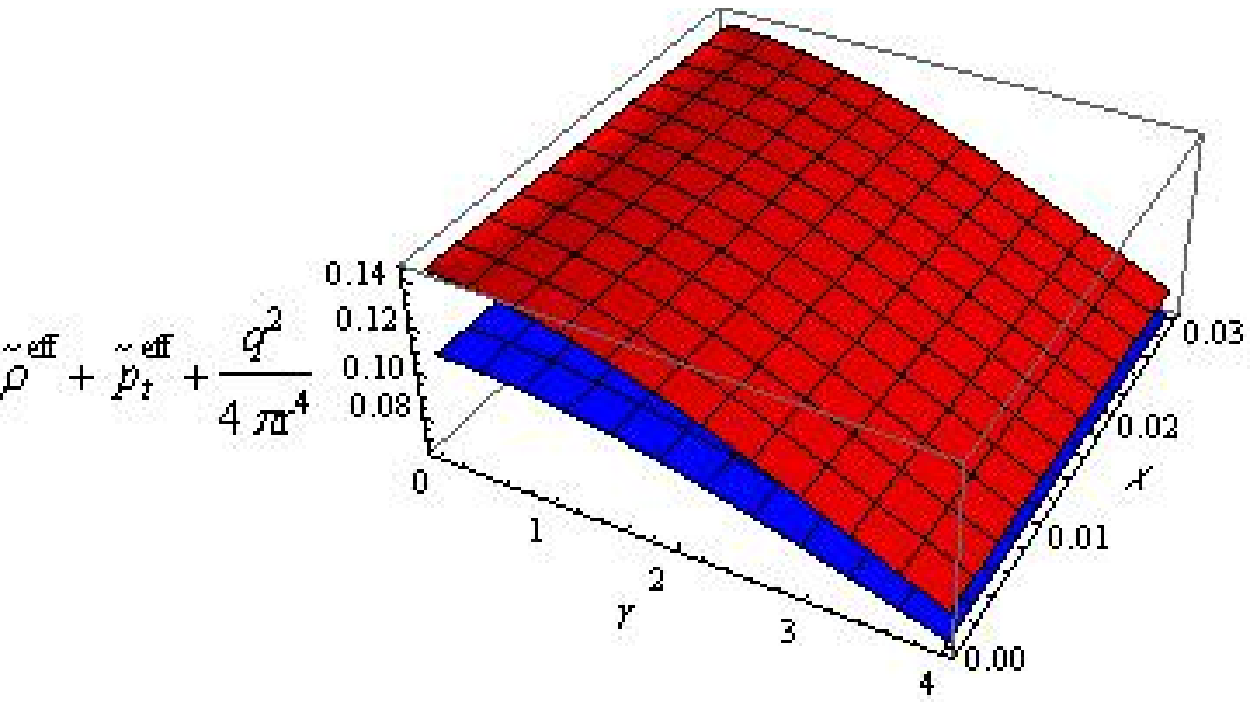,width=0.48\linewidth}
\epsfig{file=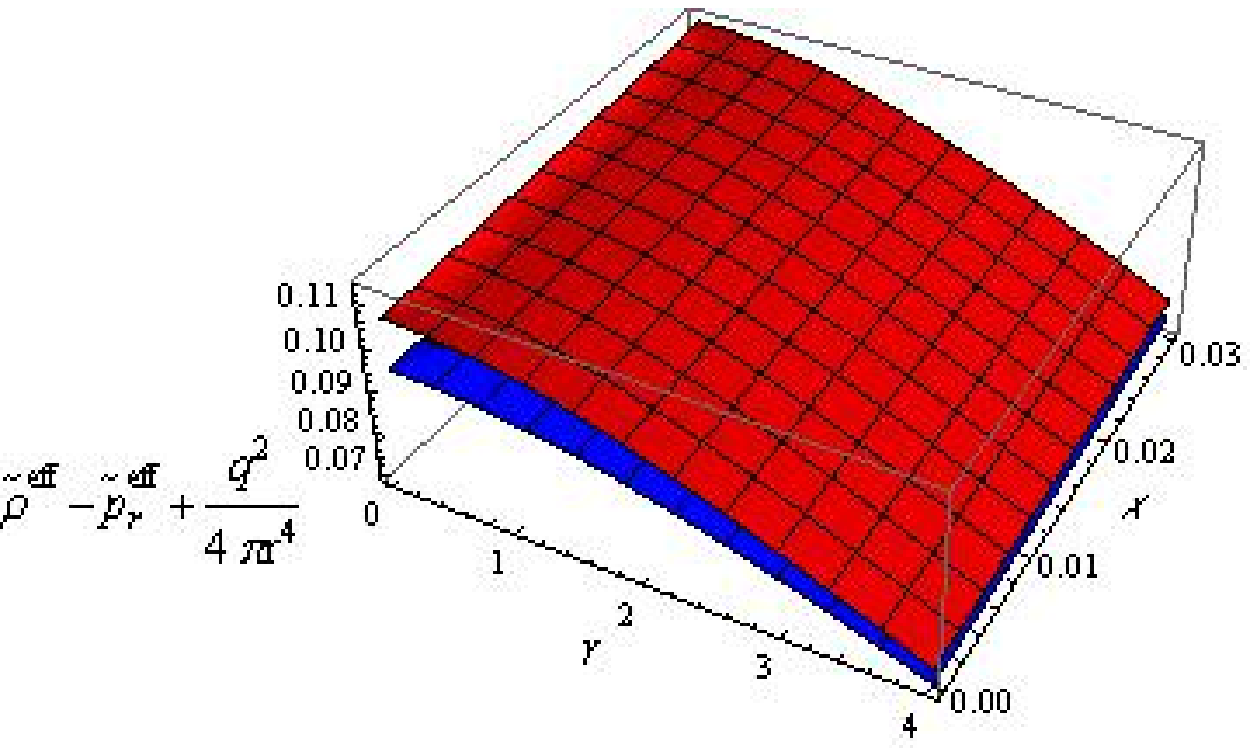,width=0.5\linewidth}
\epsfig{file=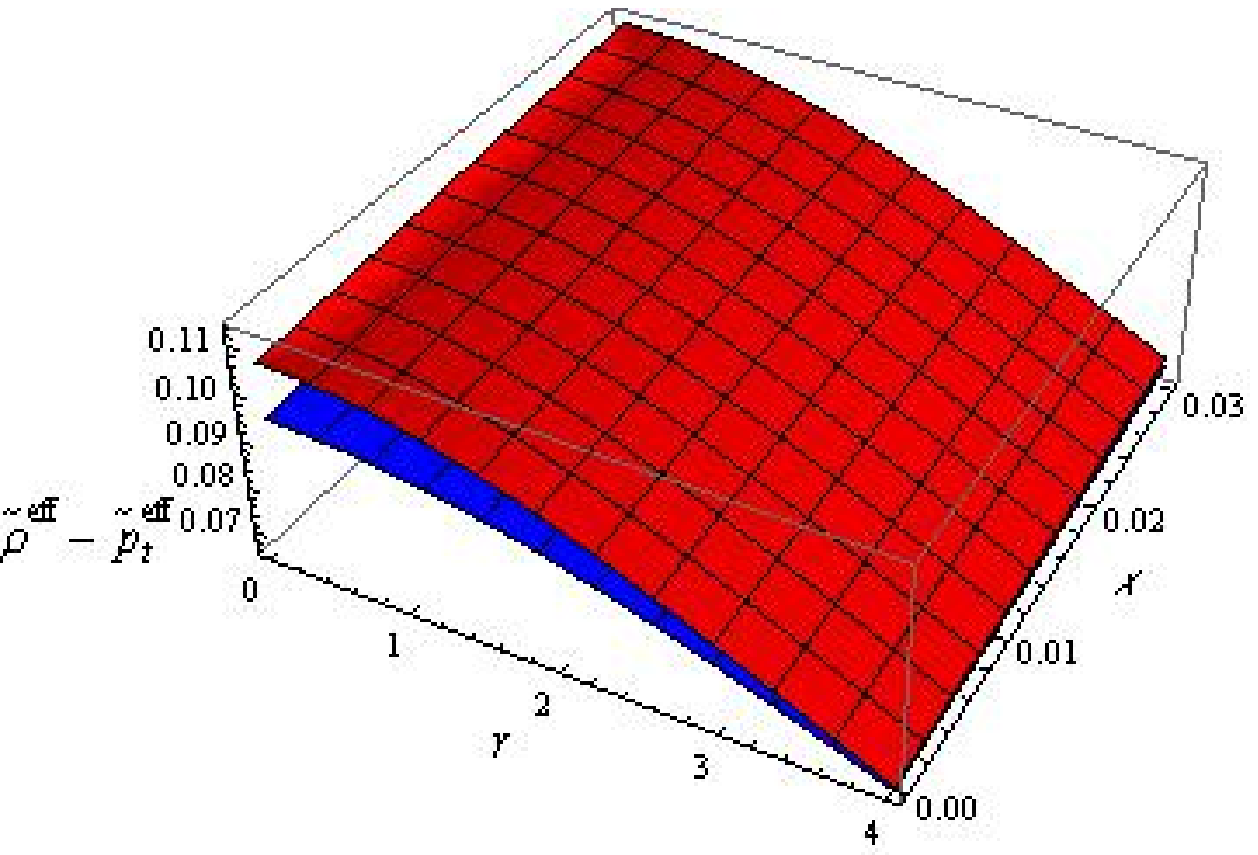,width=0.43\linewidth}
\epsfig{file=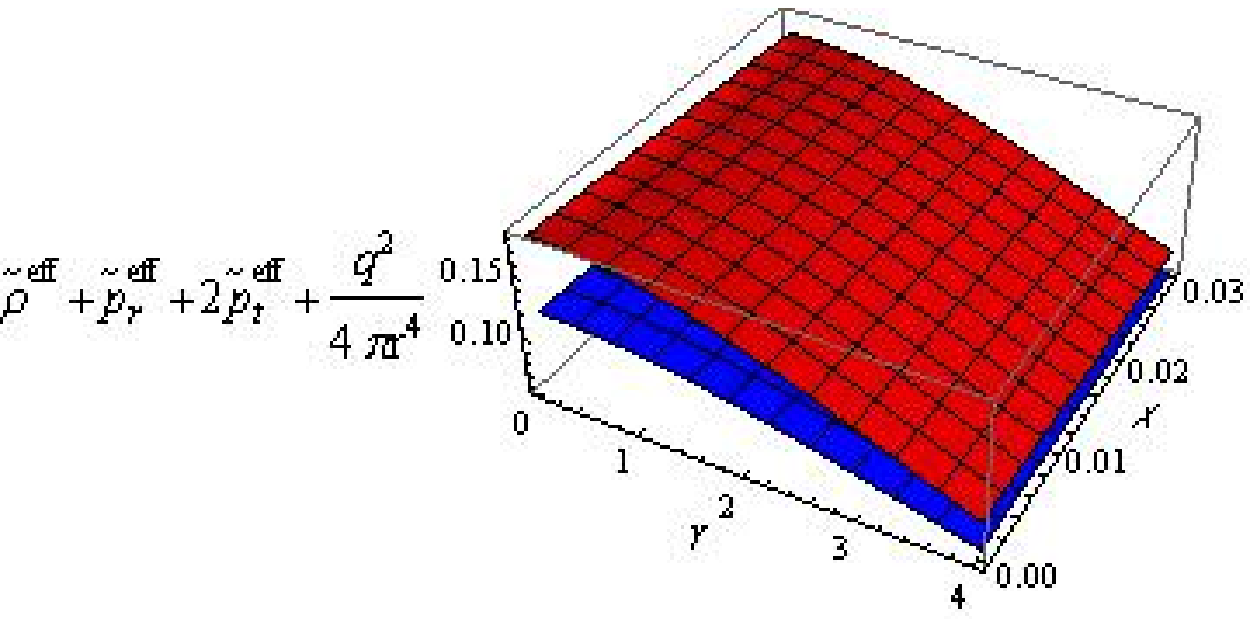,width=0.52\linewidth}\\
\caption{Behavior of energy conditions versus $r$ and $\chi$ with
$\sigma=0.2$, $Q_{0}=0.1$ (red), $Q_{0}=1$ (blue),
$M_{0}=1M_{\odot}$ and $\mathcal{R}=4M_{\odot}$ for the second
solution.}
\end{figure}
In order to investigate the behavior of second solution, we take
$\sigma=0.2$ and fix the constant $\mathcal{B}$ which is calculated
from Eqs.(\ref{49}) and (\ref{64}) whereas $\mathcal{A}$ is a free
parameter which will be considered from Eq.(\ref{47}). The graphical
behavior of $\tilde{\rho}^{eff}$, $\tilde{p}_{r}^{eff}$,
$\tilde{p}_{t}^{eff}$ and $\tilde{\Delta}^{eff}$ in the presence of
charge is represented in Figure \textbf{4}. It is observed that for
the second solution, all physical quantities (effective energy
density, effective radial and tangential pressure) decrease with the
increasing values of $Q_{0}$. The role of these matter variables is
positive, finite, regular and consistent within the interior of
stellar object. It is also found that this solution depicts
physically viable behavior only for small values of $\chi$ as
compared to the first solution. For larger values of $\chi$, this
solution does not show physically acceptable behavior.

\begin{figure}\center
\epsfig{file=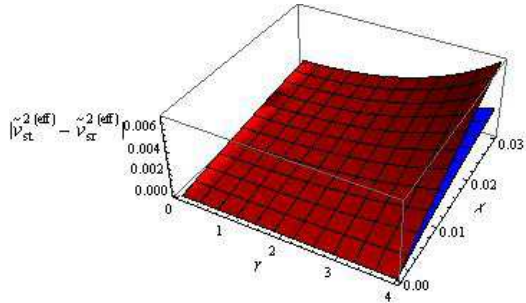,width=0.5\linewidth}\\
\caption{Plot of $|\tilde{v}^{2(eff)}_{st}-\tilde{v}^{2(eff)}_{sr}|$
versus $r$ and $\chi$ with $\sigma=0.2$, $Q_{0}=0.1$ (red),
$Q_{0}=1$ (blue), $M_{0}=1M_{\odot}$ and $\mathcal{R}=4M_{\odot}$
for the second solution.}
\end{figure}

The effect of anisotropic parameter is found to be negative for the
second solution which indicates the less massive distribution in the
interior of stellar object. The consistency of energy conditions is
shown in Figure \textbf{5} which represents that all energy
conditions are satisfied for this solution. Our second solution
reveals the potentially stable structure of compact stars as
presented in Figure \textbf{6}. It is also observed that the
stability of stellar system decreases for the larger value of charge
parameter.

\section{Concluding Remarks}

In the analysis of stellar system, the quest for new spherical
solutions has captured thoughts of many researchers. Recently, the
gravitational decoupling through MGD method has gained much
attention in obtaining the new exact solutions of self-gravitating
objects. This technique is implemented to extend the interior
isotropic spherical solutions by including the effects of
anisotropic gravitational sources. In this paper, we have used this
technique in Starobinsky form of $f(R)$ gravity with the inclusion
of charge to extend isotropic interior solution by adding the
contribution of anisotropic solution comprised in gravitational
source. In this regard, we have included a new source in charged
isotropic as well as effective energy-momentum tensor which provides
the $f(R)$ field equations corresponding to anisotropic matter
configuration.

In order to attain anisotropic solutions, we have assumed the
well-known charged isotropic Krori-Barua solution in which the
unknown constants are calculated via matching conditions. For
anisotropic solutions, we have imposed constraints on effective
pressure and effective energy density in the presence of charge
which yield the first and second solution, respectively. The
physical viability of these solutions is examined through the
graphical analysis of matter variables, effective anisotropic
factor, energy conditions and potential stability corresponding to
some specific values of charge as well as model parameter. We have
observed that both solutions are physically acceptable and show
stable structure of charged stellar object in $f(R)$ gravity.
Moreover, we have analyzed that increase in charge parameter
enhances the stability of the first solution but decreases the
stability of second solution.

Ovalle et al. \cite{26} constructed new anisotropic uncharged
spherical solutions using Tolman IV solution as interior solution
but the stable structure as well as energy conditions are not
analyzed for their solutions. Sharif and Sadiq \cite{27} employed
the Krori-Barua solution for charged spherical stellar object and
observed that only the first solution, i.e., the pressure like
constraint represents stable structure whereas the second solution,
i.e., the density like constraint disobeys the physical
acceptability. Recently, Sharif and Saba \cite{41} examined the
uncharged anisotropic spherical solutions by MGD approach in the
context of $f(\mathcal{G})$ gravity using Krori-Barua solution as
interior isotropic solution. They deduced that the stability exists
only for pressure like constraint solution. We would like to mention
here that our both solutions along with the influence of charge show
viable behavior and satisfy the required range of squared speed of
sound in $f(R)$ gravity. These solutions reduce to the solutions
obtained for uncharged case in the same gravity for $q=0$ \cite{42}.
We conclude that the $f(R)$ theory with the inclusion of charge
provides more stable distribution of stellar system as compared to
GR and $f(\mathcal{G})$ gravity.

\vspace{.25cm}

{\bf Acknowledgment}

\vspace{0.25cm}

One of us (AW) would like to thank the Higher Education Commission,
Islamabad, Pakistan for its financial support through the {\it
Indigenous Ph.D. 5000 Fellowship Program Phase-II, Batch-III.}

\end{document}